\documentclass[english,10pt]{article}
\usepackage[margin=1.2 in]{geometry}
\usepackage{float}
\usepackage{verbatim}
\usepackage{pgf}
\usepackage{mathrsfs}
\usepackage{amsmath}
\usepackage{authblk}
\usepackage{amssymb}
\usepackage{amsthm}
\usepackage{amsfonts}
\usepackage{upgreek}
\usepackage{comment}
\usepackage{algorithm}
\usepackage{algorithmic}
\usepackage{bm}
\usepackage{fancyhdr}
\usepackage{epsfig}
\usepackage{enumerate}
\usepackage{subcaption}
\usepackage{setspace}
\usepackage{mathtools}
\usepackage{tikz}
\usepackage{tikz-3dplot} 
\usepackage{pgfplots}
\usepackage{caption}
\usepackage{color}
\usepackage{dcolumn}
\pgfplotsset{compat=1.11}
\usetikzlibrary{arrows}
\usetikzlibrary{bayesnet}
\usepackage{etoolbox}

\usepackage[final]{hyperref} 
\usepackage{setspace}

\usepackage[]{biblatex}
 
\hypersetup{
	colorlinks=true,       
	linkcolor=blue,        
	citecolor=blue,        
	filecolor=magenta,     
	urlcolor=blue         
}

\title{Latent Space Network Modelling with Hyperbolic and Spherical Geometries.} 
\author[1]{Marios Papamichalis}
\author[2]{Kathryn Turnbull}
\author[2]{Simon Lunag\'{o}mez} 
\author[1]{Edoardo Airoldi}
\affil[1]{Fox School of Business, Temple University, Philadelphia, PA 19122}
\affil[2]{Department of Mathematics and Statistics, Lancaster University, Lancaster, LA1 4YW}

\begin{document}
\maketitle
\begin{abstract}

A rich class of network models associate each node with a low-dimensional latent coordinate that controls the propensity for connections to form. Models of this type are well established in the network analysis literature, where it is typical to assume that the underlying geometry is Euclidean. Recent work has explored the consequences of this choice and has motivated the study of models which rely on non-Euclidean latent geometries, with a primary focus on spherical and hyperbolic geometry. In this paper, we examine to what extent latent features can be inferred from the observable links in the network, considering network models which rely on spherical and hyperbolic geometries. For each geometry, we describe a latent space network model, detail constraints on the latent coordinates which remove the well-known identifiability issues, and present Bayesian estimation schemes. Thus, we develop computational procedures to perform inference for network models in which the properties of the underlying geometry play a vital role. Finally, we assess the validity of these models on real data. \\

Keywords: statistical network analysis, latent space network modelling, non-Euclidean geometry, Bayesian statistics.
\end{abstract}




\section{Introduction} \label{sec:intro}

The analysis of network data describing interactions among a population of interest is motivated from a diverse range of applications, and the relevance of such analysis is reflected in a vast literature (\cite{kolaczyk2009}, \cite{goldenberg2010}, \cite{saltertownshend2012}, \cite{kolaczyk2017}). A key component of this literature involves the study of networks via models which are designed to capture particular properties of interest such as heterogeneity, transitivity or power-law degree distributions. In this article, we focus on latent space network models which express the connection probabilities as a function of low-dimensional latent coordinates associated with the nodes. We pay particular attention to the choice of underlying geometry and practical challenges associated with this. \\

In the latent space approach, as introduced in \cite{hoff2002}, it is typical to impose that interactions are more likely to occur among node pairs whose latent coordinates are closer together in the latent space, where closeness is usually measured by either Euclidean distance or the dot-product. This construction offers an intuitive visualisation of the network via the latent representation and imposes desirable properties on interactions patterns, such as transitivity. A rich literature surrounds this idea which includes a range of application areas (see \cite{sweet2011modeling}, \cite{ng2018modeling}, \cite{wilson2020} for examples). Properties of Euclidean latent space network models are well understood (see \cite{rastelli2016}) and several important extensions have also been proposed, where examples include modelling community structures (\cite{handcock2007model}, \cite{krivitsky2009}, \cite{fosdick2016}), dynamic interactions (\cite{durante2014nonparametric}, \cite{sewell2015latent}, \cite{kim2018review}), multiple views (\cite{salter2017latent}, \cite{gollini2016joint}, \cite{dangelo2020}), and other non-standard network data types (\cite{mccormick2015}, \cite{sewell2019}). \\



In recent years, there has been a growing interest in modelling the latent coordinates in non-Euclidean geometries, with a primary focus on hyperbolic and spherical geometry. Modelling latent coordinates using hyperbolic space was first considered in \cite{krioukov2010}, where the authors demonstrate that this gives rise to networks with power-law degree distributions without imposing additional structure on the model as is required in the Euclidean setting (for example see \cite{krivitsky2009}). Hyperbolic latent spaces have proven useful in the context of link prediction (\cite{kitsak2020}) and network comparison (\cite{asta2014geometric}), and details for modelling in higher-dimensional hyperbolic space have been presented in \cite{kitsak2020b}. Properties beyond the degree distribution, such as clustering and the giant component, have also been explored for the model of \cite{krioukov2010} (for example, see \cite{gugelmann2012}, \cite{bode2013}, \cite{friedrich2015}, \cite{candellero2016}, \cite{kiwi2018}). As highlighted in \cite{hoff2002}, there is a clear connection between models based on dot-products in Euclidean space and models with spherical latent coordinates, and examples of latent space models with spherical coordinates can be found in \cite{young2008} and \cite{mccormick2015}. Finally, \cite{smith2019} discuss the implications of the latent geometry and compare networks with Euclidean, spherical and hyperbolic space, and \cite{lubold2020identifying} develop a hypothesis test to determine the most appropriate latent geometry for an observed network. \\

In this paper, we focus our attention on latent space network models in which the latent coordinates are represented in hyperbolic and spherical geometries. For each geometry, we specify a latent distance model in the style of \cite{hoff2002} which imposes that the latent coordinates follow a non-Euclidean Gaussian distribution and that interactions are more likely to occur between node pairs whose latent distance is small. The first assumption encodes the intuition that nodes with high-degree are likely to connect with other nodes of high-degree, and the second implies an intuitive latent space since connected nodes lie closer together. Our work is motivated by the discussion in \cite{smith2019}, who explore the relationship between the latent geometry and network properties, and we aim to address key practical considerations that are necessary for the application of these models. Firstly, since the connection probabilities are typically expressed as a function of the distance between latent coordinates, it is well understood that the latent representation suffers from non-identifiability. We characterise this source of non-identifiability and present a procedure for removing it within each geometry. Secondly, we consider Bayesian estimation for each non-Euclidean latent space model via MCMC and variational methods. Due to the challenges associated with non-Euclidean spaces, we rely on black box variational inference of \cite{ranganath2014}. \\

The remainder of this paper proceeds as follows. Section \ref{sec:non_euc_geom} provides background information for the non-Euclidean geometries relied upon in later sections and Section \ref{sec:lsnet_model} details a generic latent space model. Section \ref{sec:config_space} characterises non-identifiability in each of our non-Euclidean geometries, and presents a procedure to address this. Estimation strategies are then discussed in Section \ref{sec:estimation}, real-world examples are given in Section \ref{sec:examples} and we conclude with a discussion in Section \ref{sec:disc}. 

\section{Background: non-Euclidean Geometries}\label{sec:non_euc_geom}

This section aims to present the necessary background information for our three geometries of interest, namely hyperbolic and spherical. For each geometry, we provide an intuition, describe a distance measure and, for reasons discussed later in Section \ref{sec:lsnet_model}, describe a distribution analogous to a Normal distribution. 

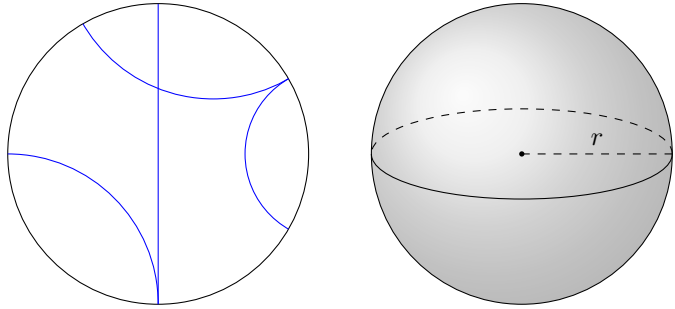
\begin{figure} 
\begin{center}
\newcommand{\hgline}[2]{
\pgfmathsetmacro{\thetaone}{#1}
\pgfmathsetmacro{\thetatwo}{#2}
\pgfmathsetmacro{\theta}{(\thetaone+\thetatwo)/2}
\pgfmathsetmacro{\phi}{abs(\thetaone-\thetatwo)/2}
\pgfmathsetmacro{\close}{less(abs(\phi-90),0.0001)}
\ifdim \close pt = 1pt
    \draw[blue] (\theta+180:1) -- (\theta:1);
\else
    \pgfmathsetmacro{\R}{tan(\phi)}
    \pgfmathsetmacro{\distance}{sqrt(1+\R^2)}
    \draw[blue] (\theta:\distance) circle (\R);
\fi
}

\begin{tikzpicture}[scale=2]
\draw (0,0) circle (1);
\clip (0,0) circle (1);
\hgline{30}{-30}
\hgline{180}{270}
\hgline{30}{120}
\hgline{0}{180}

\end{tikzpicture}
\qquad 
\begin{tikzpicture}
  \shade[ball color = gray!40, opacity = 0.4] (0,0) circle (2cm);
  \draw (0,0) circle (2cm);
  \draw (-2,0) arc (180:360:2 and 0.6);
  \draw[dashed] (2,0) arc (0:180:2 and 0.6);
  \fill[fill=black] (0,0) circle (1pt);
  \draw[dashed] (0,0 ) -- node[above]{$r$} (2,0);
\end{tikzpicture}
\caption{ Left: Poincare Disk model of hyperbolic geometry. Right: Sphere with radius $r$.} \label{fig:geometries}
\end{center}

\end{figure}

\subsection{Hyperbolic Geometry}
\label{sec:hyp_geom}


Hyperbolic geometry arises by relaxing Euclid's parallel postulate so that there exists an infinite number of parallel lines passing through a single point. This geometry is characterised by negative curvature and there exist several models for hyperbolic geometry. We focus on the Poincar\'{e} disk model for convenience and note that our network model can be described equivalently in other representations of hyperbolic geometry, such as those detailed in \cite{iversen1992}. \\

The Poincar\'{e} disk, which we denote by $\mathbb{H}$, is represented by coordinates $z \in \{ \mathbb{R}^d | \: \| z \| < 1\}$, equipped with the metric
\begin{align}
    d_{\mathbb{H}} (z_1, z_2 ) = \cosh^{-1} \left( 1 + \dfrac{2 \| z_1 - z_2 \|^2 }{(1 - \| z_1 \|^2)(1 - \| z_2 \|^2)} \right), \label{eq:poincare_dist} 
\end{align}
where $\| \cdot \|$ denotes Euclidean norm. In this model, lines are represented as either circular arcs that are orthogonal to the boundary or diameters of the disk (see left panel of Figure \ref{fig:geometries}). There exist several approaches for describing a Normal distribution in this model, such as those given in \cite{nagano2019, mathieu2019}, and we rely on the maximum entropy Normal (see \cite{pennec2006intrinsic,said2014new,hauberg2018directional}). This distribution is given by
\begin{align}
\mathcal{N}_{\mathbb{H}} \left( z \mid {\mu}, \sigma^2 \right)= \frac{1}{Z(\sigma)} \exp\left(-\frac{d_{\mathbb{H}}({\mu}, z)^2}{2\sigma^2}\right) \label{eq:hyp_gauss}
\end{align}
where $\sigma >$ 0 is a dispersion parameter, ${\mu} \in \mathbb{H}$ is the mean, and $Z(\sigma)$ is the normalising constant. For $d=2$ we have
\begin{align}
    Z( \sigma ) = 2 \pi \sqrt{\dfrac{\pi}{2}} \sigma e^{\sigma^2 / 2} \mbox{erf}\left( \dfrac{ \sigma }{ \sqrt{2} } \right)
\end{align}
Since it is not possible to obtain samples directly from this distribution, we rely on th rejection sampler a detailed in \cite{mathieu2019}. 

\subsection{Spherical Geometry}
\label{sec:ell_geom}

Spherical geometry is concerned with points lying on the unit sphere $\mathbb{S}^d = \left\{ z \in \mathbb{R}^{d+1} | \| z \| = 1 \right\}$. This geometry is characterised by positive curvature, and is closely related to elliptic geometry. For two points $z_1, z_2 \in \mathbb{S}^d$ the spherical distance is given by the angle between vectors from the origin to each of $z_1$ and $z_2$, namely
\begin{align}
  d_{\mathbb{S}}(z_1, z_2) = \cos^{-1} \left( z_1^T z_2 \right). \label{eq:dist_sphere}
\end{align}

Several distributions which mimic properties of a Normal have been developed for the sphere for directional statistics (see \cite{paine2018}, \cite{kent1982}, \cite{mardia2008}). We focus on the von-Mises-Fisher distribution, whose density function is given by
\begin{align}
  f_{\mathbb{S}^d}( z ;{ \mu },\kappa )= \frac{\kappa ^{d/2-1}}{(2\pi )^{d/2}I_{d/2-1}(\kappa )} \exp \left( \kappa {{\mu }}^{T}\mathbf {z} \right), \label{eq:vmf}
\end{align}
where $d$ is the dimension of $\mathbb{S}^d$, $\kappa \geq 0$ is the concentration parameter, $\mu$ represents the mean and $I_v$ denotes the modified Bessel function of the first kind. We focus on the setting where $d=3$, where the normalising constant simplifies to give
\begin{align}
f_{\mathbb{S}^s}( z ;{ \mu },\kappa )= \dfrac{\kappa}{2 \pi (e^{\kappa} - e^{-\kappa} )} \exp \left({\kappa {{\mu }}^{T}\mathbf {z} }\right),  
\end{align}

For this distribution, a larger value of $\kappa$ implies that the distribution is more concentrated around the mean direction $\mu$. This distribution is unimodal for $\kappa >0$ and uniform on the sphere for $\kappa =0$. Samples from this distribution can be obtained via the rejection sampler described in \cite{wood1994simulation} and implemented in the R package \texttt{movMF} \cite{movmf}. 

\section{Non-Euclidean Latent Space Network Modelling}\label{sec:lsnet_model}

In this latent space network model of \cite{hoff2002}, nodes of a network are associated with low-dimensional latent coordinates which capture their tenancy to form ties. In this section we outline a generic model which we then use to detail a latent space network model for each of the geometries outlined in Section \ref{sec:non_euc_geom}. Throughout we let $\mathbb{H}^2 and \mathbb{S}^2$ denote hyperbolic and spherical geometry in two dimensions, respectively. 

\subsection{Generic Model and Notation}
\label{sec:gen_mod}

Consider a network on $N$ nodes, indexed by $[N] = \{1,2,\dots,N\}$. We let $\mathcal{Y} = ( y_{ij} )_{i,j,\in [N]}$ denote the $N \times N$ adjacency matrix with binary entries $y_{ij} \in \{0,1\}$, where $y_{ij}=1$ if there is an edge between nodes $i$ and $j$ and $y_{ij}=0$ otherwise. We assume that there are no self ties, so that $y_{ii}=0$ for $i \in [N]$ and that connections are symmetric, so that $y_{ij}=y_{ji}$. \\

In this model, each node is assigned a $d$-dimensional latent coordinate which describes the probabilities of connections forming. We let $z_i$ denote the latent coordinate of the $i^{th}$ node, and $\bm{Z} = \{ z_i \}_{i=1}^N$ denotes the entire latent representation. \\

Similarly to \cite{smith2019}, we write a generic latent space model for all geometries as
\begin{align}
    Y_{ij} &\sim \mbox{Bernoulli}(p_{ij}) && (i,j) \in \left\{ [N] \times [N] | i < j \right\} \nonumber \\ 
    \mbox{logit} (p_{ij}) &= \alpha - d_{\mathcal{G}}(z_i, z_j)   \label{eq:generic_model}\\
    z_i &\sim f_{\mathcal{G}}(z | \theta_z) && i \in [N] \nonumber
\end{align}
where $d_{\mathcal{G}}(z_i, z_j)$ represents the distance measure between coordinates $z_i$ and $z_j$ in the geometry $\mathcal{G} \in \{\mathbb{H}, \mathbb{S}\}$, $\theta_z$ denotes additional parameters which define the distribution on $\bm{Z}$ and $\alpha$ describes the base-rate tendency for connections to form. This model imposes that nodes whose latent coordinates are close in terms of $d_{\mathcal{G}}(z_i, z_j)$ are more likely to be connected and, since the distance is a metric, the triangle inequality implies transitive relationships, in which ``friends of friends are also friends'', are likely.  We note here that it is straightforward to adapt the above model to express non-binary and asymmetric tries. \\

\begin{algorithm}[t]
\begin{algorithmic}
\STATE Sample $Z = \{z_i\}_{i=1}^N$ such that $z_i \overset{\text{iid}}{\sim} f_{\mathcal{G}}(z \mid \theta_z)$, for $i \in [N]$.
\STATE For $i = 1, 2,\dots, N-1$
\STATE \hspace{.5cm} For $j = i+1, \dots, N$
\STATE \hspace{1cm} Calculate $p_{ij} = 1/(1 + \exp\{ - (\alpha - d_{\mathcal{G}}(z_i, z_j) ) \} ) $.
\STATE \hspace{1cm} Sample $y_{ij}$ from $\mbox{Bernoulli}(p_{ij})$ 
\STATE \hspace{.5cm} End 
\STATE End 
\end{algorithmic}
\caption{Sample a graph according to \eqref{eq:generic_model}} \label{alg:sample_gen_graph}
\end{algorithm}

It is clear that the choice of distribution $f_{\mathcal{G}}(z \mid \theta_z)$ will impact the properties of networks generated according to \eqref{eq:generic_model}. In the Euclidean setting we may, for example,  assume that the latent coordinates follow a $d$-dimensional Normal distribution $\mathcal{N}_d(\mu, \Sigma)$. This choice imposes that nodes which are positioned close to the mode $\mu$ will share a larger number of connections than those positioned further from $\mu$, and that nodes with high degree are likely to be connected to other nodes of high degree. For each of the geometries discussed in Section \ref{sec:non_euc_geom}, it is common to take uniformly distributed coordinates (see \cite{krioukov2010} and \cite{smith2019}) and a distribution analogous to the Euclidean Normal has yet to be considered in this context. \\

Algorithm \ref{alg:sample_gen_graph} describes a generic procedure for sampling a graph from this model given a choice of geometry, $N, \alpha$ and $\theta_z$. Since each connection is modelled independently given the latent representation, we can express the likelihood of observing $\mathcal{Y}$ conditional on $\bm{Z}$ and $\alpha$ as
\begin{align}
    p(\mathcal{Y} | \bm{Z}, \alpha) = \prod_{i<j} p_{ij}^{y_{ij}} (1-p_{ij})^{1-y_{ij}}. \label{eq:py_given_u}
\end{align}
Given specification of prior distributions for the parameters $\alpha$ and $\theta_z$, we may obtain posterior samples via Bayesian estimation procedures. This will be discussed further in Section \ref{sec:estimation}. 


\section{Non-identifiability of the Latent Coordinates}
\label{sec:config_space}

It is well understood that latent space network models suffer from non-identifiability of the latent coordinates. To see this, note that, in \eqref{eq:generic_model}, the connection probabilities are modelled as a function of $\bm{Z}$ so that $p_{ij}$ will remain constant under transformations of $\bm{Z}$ which preserve $d_{\mathcal{G}}(z_i,z_j)$. In the Euclidean case, these transformations are given by compositions of translations, reflections and rotations. \\

This source of non-identifiability is typically addressed via a Procrustes transformation in which the estimates of the coordinates are mapped onto a pre-specified set of reference coordinates (for example, see \cite{hoff2002}). Whilst extensions of this method have been considered in non-Euclidean settings \cite{lin2021hyperbolic,tabaghi2021procrustes,goodall1991procrustes,bingham1992approximating}, we instead take a similar approach to \cite{turnbull2019} who avoid this post-processing step by appropriately constraining a subset of the latent coordinates (see also \cite{mccormick2015}). This draws on the notion of Bookstein coordinates from shape theory (see \cite{dryden1998}) and avoids an additional calculation for each interation of the inferential procedures discussed in Section \ref{sec:estimation}. In this section, we will describe the set of distance-preserving transformations for each of the geometries discussed in Section \ref{sec:non_euc_geom} and use this to characterise constraints on $\bm{Z}$ which ensure identifiability. We stress that our discussion focuses on the case when $d=2$, but the concepts may be extended to higher-dimensions.

\subsection{Generic Procedure}
\label{sec:config_generic}

Given an expression for the distance-preserving transformations, namely the isometries, we may map the latent representation onto an analogue of Bookstein coordinates (see \cite{dryden1998}). Since we restrict to the case where $d=2$ and $p_{ij}$ is expressed in terms of distances, we must fix one coordinate and constrain two further coordinates to remove all sources of non-identifiability for each of the geometries. Let $I_{\theta_I}(z)$ denote the class of isometries and $z_1^{*}, z_2^{*}$ and $z_3^*$ denote the constrained coordinates, hereon referred to as \emph{anchor coordinates}. A generic procedure proceeds as follows, and we discuss details for each geometry in the follow subsection.

\begin{enumerate}
\item Given an initial latent representation $\bm{Z}$, choose indices for the anchor coordinates, denoted $\{i_1, i_2, i_3\}$.
\item Determine the isometry which satisfies $z_{i_j} \mapsto z_{i_j}^*$ for $j = 1,2,3$. This corresponds to a particular instance of $\theta_I$.
\item Take $\bm{Z}^* = I_{\theta_I}(\bm{Z})$ and, throughout the estimation procedure, keep $z_{i_1}^*$ fixed and appropriately constrain $z_{i_2}^*$ and $z_{i_3}^*$.
\end{enumerate}

To understand why we require three anchor coordinates, consider Euclidean isometries. In this setting, the first anchor point removes the effect of translations, the second anchor point removes the effect of rotations and the third anchor point removes the effect of reflections in the $x$-axis. Note that strictly fixing the two latent coordinates, such as in \cite{turnbull2019}, is too restrictive in this setting since this will imply a constant value of $d_{\mathcal{G}}(z_{i_1}^*, z_{i_2}^*)$ and this distinction is due to differences in the model specification. Finally, we note that if $d>2$, we will require additional points to be constrained for each geometry.

\subsection{Configuration Space for Each Geometry}

We now discuss the details of the procedure outlined in Section \ref{sec:config_generic} for each of the non-Euclidean geometries discussed in Section \ref{sec:non_euc_geom}. 

\subsubsection{Hyperbolic Geometry}

The Poincar\'{e} disk may also be represented by complex coordinates in the unit disk $\mathbb{D} = \{ z \in \mathbb{C} | |z| < 1\}$. With this representation, isometries are given by transformations $h$ of the form
\begin{align}
  h(z) = \beta \dfrac{z - z_0}{1 - \bar{z}_0 z}, \label{eq:poincare_isom}
\end{align}
where $| \beta | = 1$ is some angle and $z_0 \in \mathbb{D}$. This result is well known and can be found in, for example, \cite{hvidsten2016}. \\

These transformations can be viewed as compositions of rotations about the centre of the Poincar\'{e} disk and inversions through circles that are perpendicular to the boundary of the disk, followed by reflections in the vertical axis. We choose our first anchor point to be the origin and our second anchor point to be constrained to lie on the positive $x$-axis, so that
\begin{align}
    z_{i_1}^*  = 0 + 0i, \hspace{1cm} z_{i_2}^* = a + 0i \label{eq:poincare_anchor}
\end{align}
where $a > 0$. To apply this isometry, we first need to determine the value of $a$ which preserves the distance between $z_{i_1}, z_{i_2}$ and $z_{i_1}^*, z_{i_2}^*$. Then, given this, we can determine appropriate values of $z_0$ and $\beta$. Straightforward calculations return
\begin{align}
  a = \sqrt{ \dfrac{\mbox{cosh} d(z_{i_1}, z_{i_2}) - 1}{1 + \mbox{cosh}d(z_{i_1}, z_{i_2})} }, \hspace{1cm}
    z_0 = z_{i_1}, \hspace{.5cm} \mbox{ and } \hspace{.5cm}
    \beta = \sqrt{ \dfrac{\mbox{cosh} d(z_{i_1}, z_{i_2}) - 1}{1 + \mbox{cosh}d(z_{i_1}, z_{i_2})} } \left( \dfrac{\bar{z}_{i_1}z_{i_2} - 1}{z_{i_2} - z_{i_1}} \right).
\end{align}
For eliminating reflections in the first axis we ensure that the third anchor point has positive imaginary part, by reflecting the latent positions along the horizontal axis, if necessary. Details of these calculations can be found in Appendix \ref{app:config_hyp}. \\

We may now learn the latent representation on the new coordinate set given by $\bm{Z}^* = \{ h(z_i) \}_{i \in [N]}$, where coordinates $i_1$ and $i_2$ are constrained according to \eqref{eq:poincare_anchor}. Furthermore, suppose that we have $\bm{Z}$ which are distributed according to a hyperbolic Normal with parameters $\bm{\mu}$ and $\sigma$, so that $z_i \sim \mathcal{N}_{\mathbb{H}}(\mu, \sigma)$ for $i \in [N]$. It follows that, after applying the isometry $h$ to obtain $\bm{Z}^*$, we have $z_i^* \sim \mathcal{N}_{\mathbb{H}}\left( h(\mu), \sigma \right)$ for $i \in [N]$.

\subsubsection{Spherical geometry}
\label{sec:config_spherical}

The distance-preserving transformations of the sphere $\mathbb{S}^2$ can be viewed as a composition of rotations about each of the three axes. For $z \in \mathbb{S}^2$, we write $\bm{z} = (z_1, z_2, z_3)$ and let $R_{z_j, \theta_{j}}$ denote a rotation of angle $\theta_{j}$ about the $z_j$ axis, for $j=1,2,3$. Expressions for these rotation matrices are well known, and are reproduced in \eqref{eq:3d_rot_mats} in Appendix \ref{app:config_sphere} for completeness. \\

We let $R_{\theta_{1}, \theta_{2}, \theta_{3}}$ represent the transformation obtained by rotating angle $\theta_{1}$ around the $z_1$ axis, followed by a rotation of angle $\theta_{2}$ around the $z_2$ axis, followed by a rotation of angle $\theta_{3}$ around the $z_3$ axes, so that $R_{\theta_{1}, \theta_{2}, \theta_{3}} =  R_{z_3, \theta_{3}} R_{z_2, \theta_{2}} R_{z_1, \theta_{1}}$. Taking advantage of the order of these rotations, we choose our first two anchor coordinates to be
\begin{align}
  \bm{z}_{i_1}^* = (0,0,1) \hspace{.5cm} \mbox{ and } \hspace{.5cm} \bm{z}_{i_2}^* = \left(a,0, b \right), \label{eq:anchor_sphere}
\end{align}
where $0<a<1$. Since we are applying an isometry, we obtain $b = \cos \left( d_{\mathbb{S}} (\bm{z}_{i_1}, \bm{z}_{i_2} ) \right)$ and $a = \sqrt{1 - b^2}$, where we take the positive root for $a$. \\

Straightforward calculations return
\begin{align}
\tan \theta_{1} &= \dfrac{ z_{i_1,2} }{ z_{i_1,3} } \\
\tan \theta_{2} &= \dfrac{- z_{i_1,1} }{ z_{i_1, 2} \sin \theta_{1} + z_{i_1,3} \cos \theta_{1}} \\
\tan \theta_{3} &= \dfrac{ z_{i_2,3} \sin \theta_{1} - z_{i_2,2} \cos \theta_{1} }{ z_{i_2,1} \cos \theta_{2} + \sin \theta_{2} ( z_{i_2,2} \sin \theta_{1} + z_{i_2,3} \cos \theta_{1} ) },
\end{align}  
where $z_{i,j}$ denotes the $j^{th}$ element of $\bm{z}_i$. Finally, we choose our third anchor coordinate $z_{i_3}^* = (c,d,e)$ where $d > 0$. If $z_{i_3}^*$ has a negative entry after the above rotation matrices have been applied, we further reflect according to $Ref_{1,3}= \mbox{diag}(1, -1, 1)$. Details of these calculations are given in Appendix \ref{app:config_calc}. \\

As in the previous section, we can now estimate the latent coordinates on $\bm{Z}^* = \{ R_{\theta_{1}, \theta_{2}, \theta_{3}} \bm{z}_i \}_{i \in [N]}$, where $z_{i_1}^*$ and $z_{i_2}^*$ are constrained according to \eqref{eq:anchor_sphere}.
Now, suppose that the coordinates $\bm{Z}$ are distributed according to a von Mises-Fisher distribution with parameters $\kappa$ and $\mu$. Since the Jacobian of the transformation given by $R_{\theta_{1}, \theta_{2}, \theta_{3}}$ is equal to 1, it follows that $\bm{Z}^*$ follow a non Mises-Fisher distribution with parameters $\kappa^* = \kappa$ and $\mu^* = R_{\theta_{1}, \theta_{2}, \theta_{3}} \mu$.

\section{Estimation}
\label{sec:estimation}

We now consider estimating the parameters of the model given in \eqref{eq:generic_model} for each geometry. To begin, we outline a generic MCMC sampler in Section \ref{sec:post_samp} which allows for asymptotically exact inference. However, since evaluation of the posterior scales poorly with $N$, we also consider estimation via variational methods in Section \ref{sec:vi} to improve scalability. In particular, we explore using Black Box variational inference for the continuous cases, namely hyperbolic and spherical geometries.

\subsection{MCMC sampler}\label{sec:post_samp}

For each model, we assume that the parameter controlling the base rate tendency of edges to be formed is a priori Normally distributed. Given this, we can express the posterior distribution for the generic model \eqref{eq:generic_model} as
\begin{align}
  p( \bm{Z}, \alpha, \theta_z | \mathcal{Y} ) \propto p( \mathcal{Y} | \bm{Z}, \alpha ) \prod_{i \in [N]} p( z_i | \theta_z ) p(\alpha | m, \sigma) p( \gamma ), \label{eq:post}
\end{align}
where $p( \mathcal{Y} | \bm{Z}, \alpha )$ is given in \eqref{eq:py_given_u}, $p(\alpha | m, \sigma ) = \mathcal{N}(m, \sigma)$, and $p( \gamma )$ denotes an additional prior distribution parameterised by hyperparameters $\gamma$. The particular form of the posterior will depend on the geometry being considered. In all cases, the posterior does not admit a closed form expression and so we use MCMC to obtain samples from the posterior. \\

We rely on a Metropolis-within-Gibbs to obtain posterior samples, and a generic outline of the MCMC scheme is presented in Algorithm \ref{alg:mcmc_scheme}. Our focus in this section is to present an intuition of the MCMC scheme, and we refer to Appendix \ref{app:mh_details} for details specific to each geometry. As discussed in Section \ref{sec:config_space}, we opt to estimate $\bm{Z}$ on a restricted space which ensures identifiability. This step is incorporated into the initialisation procedure and we index the anchor coordinates by $\{ z_{i_1}, z_{i_2}, z_{i_3} \}$ throughout. 

\begin{algorithm}[t]
  \caption{Outline of MCMC sampler} \label{alg:mcmc_scheme}
  \begin{algorithmic}
    \STATE \textbf{Input}: observations $\mathcal{Y}$, number of iterations $L_{max}$ and anchor coordinates $\{i_1, i_2, i_3\}$
    \STATE \textbf{Initialise}: determine initial values for $\bm{Z}^{(0)}, \alpha^{(0)}, m^{(0)}, \sigma^{(0)}, \theta_z^{(0)}$ (see Appendix \ref{app:mh_details}).
    \STATE Determine the isometry $I$ which takes $\{i_1,i_2, i_3\}$ to the anchor coordinates, and take $\bm{Z}^* = \{ I(z_i) \}_{i \in [N]}$
    \STATE \textbf{Iterate over update steps}:
    \STATE For $l = 1, 2, 3, \dots, L_{max}$
    \STATE \begin{enumerate}
      
    \item Determine $m^{(l)}$ and $\sigma^{(l)}$ via a MH step analogously to \eqref{eq:alph_update} and \eqref{eq:zi_upd}.
      
    \item Propose $\alpha^{(prop)}$ via a symmetric random walk and set $\alpha^{(l)} = \alpha^{(prop)}$ with probability
              \begin{align}
          AR_{\alpha} = \min \left\{ 1, \dfrac{p( \mathcal{Y} | \bm{Z}^{(l-1)}, \alpha^{(prop)} ) p( \alpha^{(prop)} | m^{(l)}, \sigma^{(l)} ) p( \gamma^{(l)} ) }{ p( \mathcal{Y} | \bm{Z}^{(l-1)}, \alpha^{(l-1)} ) p( \alpha^{(l-1)} | m^{(l)}, \sigma^{(l)} ) p( \gamma^{(l)} ) } \right\} \label{eq:alph_update}
              \end{align}
              Otherwise, set $\alpha^{(l)} = \alpha^{(l-1)}$.
    \item Let $\bm{Z}^{(l)} = \bm{Z}^{(l-1)}$ denote the current estimate of $\bm{Z}$. For $i \in [N] \backslash \{i_1\}$: \\
       Propose $z_i^{(prop)}$ via a symmetric random walk and set $\bm{Z}^{(prop)}$ to be the current state of $\bm{Z}$ with $i^{th}$ row set to $z_i^{(prop)}$. Then set $\bm{Z}^{(l)} = \bm{Z}^{(prop)}$ with probability
        \begin{align}
          AR_{z_i} = \min \left\{ 1, \dfrac{p( \mathcal{Y} | \bm{Z}^{(prop)}, \alpha^{(l)} ) p( z_i^{(prop)} | \theta_z^{(l)} )}{p( \mathcal{Y} | \bm{Z}, \alpha^{(l)} ) p( z_i | \theta_z^{(l)} )} \right\} \label{eq:zi_upd}
        \end{align}
        Note that the update for $i \in \{i_2, i_3\}$ will be restricted according to the geometry of the latent coordinates.
        
    \end{enumerate}
    End
  \end{algorithmic} 
\end{algorithm}

\subsection{Variational Bayesian Inference}\label{sec:vi}


Since evaluation of the posterior \eqref{eq:post} requires $O(N^2)$ calculations, it is well understood that estimation via MCMC scales poorly with $N$. To resolve this in the Euclidean setting, several likelihood approximations have been proposed (\cite{raftery2012}, \cite{rastelli2018}) and, for specific model forms, more efficient MCMC samplers have been developed \cite{spencer2020}. We instead opt to variational methods, which determine a computationally cheaper approximation to the posterior via optimisation. \\

Although variational methods have previously been considered in the latent space setting (\cite{salter2017latent}, \cite{gollini2016joint}), the non-Euclidean geometry offers unique challenges due to the form of the distributions on $\bm{Z}$. In `vanilla' variational inference, the optimal member of a family of distributions $\mathcal{Q}$ is determined by maximising the evidence lower-bound (ELBO) given by
\begin{align}
    \mbox{ELBO}(q) = \mathbb{E}_q \left[ p( \bm{Z}, \alpha, \mathcal{Y}) \right] - \mathbb{E}_q \left[ q( \bm{Z}, \alpha )\right], \label{eq:elbo}
\end{align}
where $q \in \mathcal{Q}$. It can be shown that maximising the ELBO is equivalent to minimising the Kullback-Leibler between the variational distribution $q$ and the target $p$ (see \cite{blei2017}). Note also that we have supressed the dependence of additional model parameters in \eqref{eq:elbo} for convenience.  We find this expression intractable and instead rely on Black Box Variational Inference (BBVI) of \cite{ranganath2014}. This scheme instead determines the variational approximation through Monte Carlo estimates of the gradient of the ELBO and avoids direct calculation of \eqref{eq:elbo}. \\

In order to use BBVI we must specify a mean-field variational family, so that each parameter in the target is assigned its own variational parameter. For the posterior \eqref{eq:post}, we take the variational family to be parameterised as
\begin{align}
  q = q( \alpha | \tilde{m}, \tilde{s} ) \prod_{i \in [N]} q(z_i | \tilde{\theta}_z ), \label{eq:q}
\end{align}
where $q( \alpha | \tilde{m}, \tilde{s} ) = \mathcal{N}(\tilde{m}, \tilde{s})$ and the specification of $q(z_i | \tilde{\theta}_z )$ is geometry dependent. Implementation of this scheme requires expressions of the derivative of $\log(q)$ with respect to each of the variational parameters $\tilde{m}, \tilde{s}$ and $\tilde{\theta}_z$. We refer to Appendix \ref{app:vi_details} for the necessary details and a description of the BBVI algorithm. Note that the initialisation step for this approach can be carried out in an analogous way to the MCMC scheme (see Appendix \ref{app:mh_details}).  \\

\section{Examples}\label{sec:examples}

To examine the performance of our proposed methodology we consider a real world data example for each of the geometries previously discussed. We estimate each model using both the MCMC and variational inference schemes detailed in Section \ref{sec:estimation}, and compare these in terms of parameter estimates and predictive distributions. In particular, we compare the posterior mean for model parameters from the MCMC with the estimates of variational means and explore the posterior predictive probabilities of a link forming for each estimation procedure. Intuitively, if these algorithms perform comparably well, we expect the posterior and variational means to be close and the predictive probabilities to show clear separation between present and absent links. We note here that a reasonable comparison can be made between estimates of the latent positions since we rely on the procedure for removing non-identifiability outlined in Section \ref{sec:config_space}. Each algorithm was implemented in R and geometry specific details are given in the Appendix. Code to implement our methodology is available at https://github.com/MariosPapamix. 

\subsection{Florentine Family}
\label{sec:florentine_family_spherical}

In this section we consider the Florentine family network dataset (see \cite{breiger1986, mueller1981}) which details marriage relations among $N=15$ families in fifteenth-century Florence. For this example, $y_{ij}=1$ indicates that there was a marriage between families $i$ and $j$ and $y_{ij}=0$ otherwise. We analyse this dataset using our model with the assumption that the latent coordinates lie in spherical geometry. To justify this choice, we compare the quality of multidimensional scaling (MDS) estimates when the underlying geometry is assumed to be spherical and Euclidean. We use graph distance as a proxy for the latent node distances and rely on the \text{smacof} R package \cite{smacofpackage} to determine the spherical MDS estimates. To compare the estimates we consider the stress which serves as a goodness-of-fit measure for MDS (for example, see \cite[Section 3]{borg2005modern}) and, using this approach, we determine that the spherical embedding is more appropriate for this data example. \\

\begin{figure}[H]
  \centering
    \hspace*{-2cm}
  \includegraphics[scale=0.8]{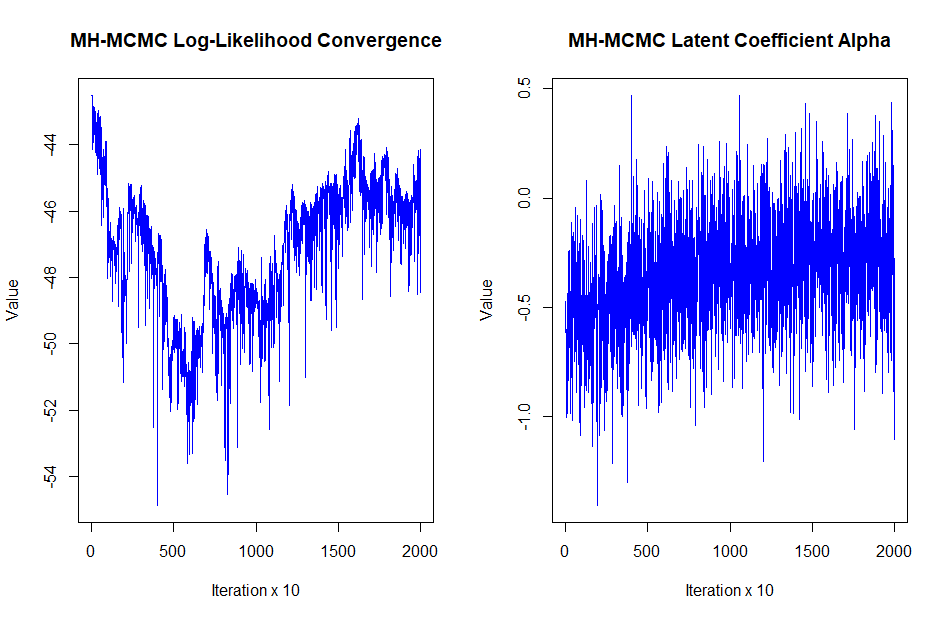}
  \caption{Thinned MCMC traceplot of log-likelihood (left) and $\alpha$ (right) for Florentine family modelled with spherical latent coordinates. For each chain we plot 2000 equidistant samples from the 20000 posterior samples. }
  \label{fig:convergence_florentine}
\end{figure}

\begin{figure}[H]
    \centering
     \hspace*{-2cm}
  \includegraphics[scale=0.8]{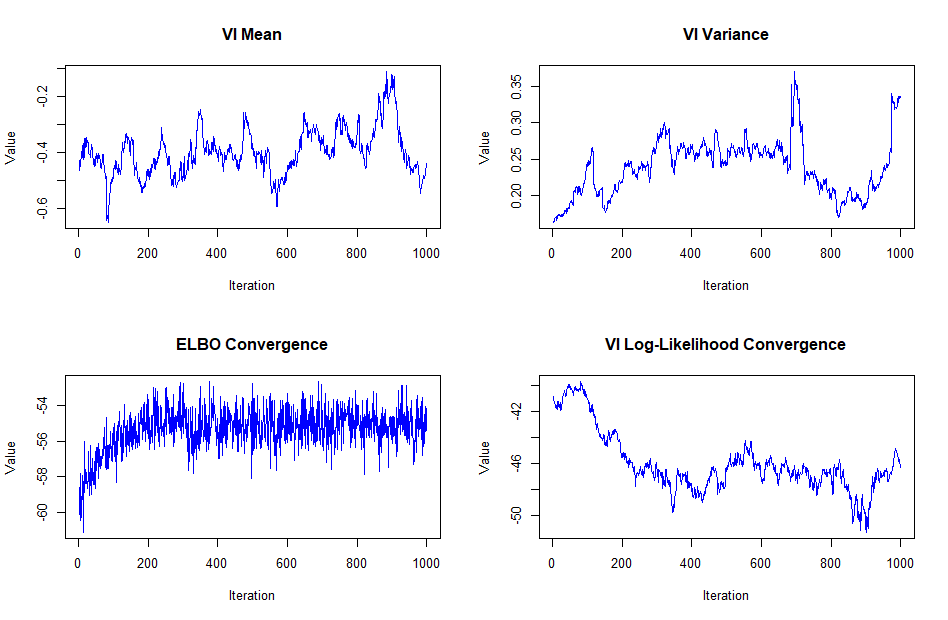}
  \caption{
  Summary of BBVI variational parameters for Florentine family modelled with spherical latent coordinates. BBVI is implemented with 1000 iterations amd $S=20$. Top, left to right: variational mean $\tilde{m}$ and variance $\tilde{\sigma}$ of $\alpha$. Bottom, left to right: ELBO and log-likelihood showing convergence for BBVI.
  }
  \label{fig:florentine_parameters}
\end{figure}

\begin{table}
\centering
\fbox{%
\begin{tabular}{| l  l  ll|}
\hline
Dataset  & Nodes & MCMC estimate & BBVI estimates
 \\
\hline            
Florentine family &15   & $\hat{\alpha}
=-0.53$&$(\tilde{m},\tilde{\sigma})=(-0.51,0.232)$ \\
\hline            
\end{tabular}}
\caption{Estimation of the base-rate parameter $\alpha$ for the Florentine family dataset when the latent geometry is assumed to be spherical. This table reports the posterior mean of $\alpha$ for the last 2000 MCMC estimations and the estimates of the variational parameters for $\alpha$.}
\label{table:alpha_florentine}
\end{table}

\begin{figure}[H]
  \centering
  
  \includegraphics[width=\textwidth]{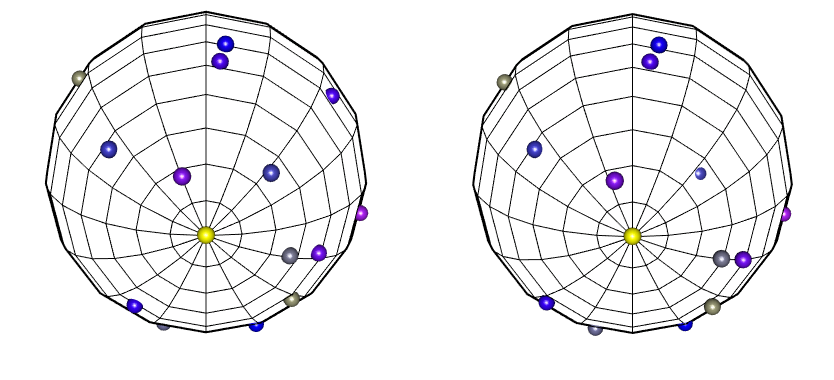}
  \caption{Summary of estimates of spherical latent positions $\bm{Z}$ for the Florentine family dataset. Left: Posterior means of $\bm{Z}$ obtained via MCMC. Right: variational mean parameter estimates $\tilde{\bm{Z}}$ obtained via BBVI. Note that we report the Fr\'{e}chet mean since this respects the spherical geometry. We also maintained the same anchor coordinates for each procedure so that the corresponding estimates are comparable. }
  \label{fig:slatent}
\end{figure}

\begin{figure}[H]
  \centering
 \hspace*{-2cm}
 \includegraphics[scale=1]{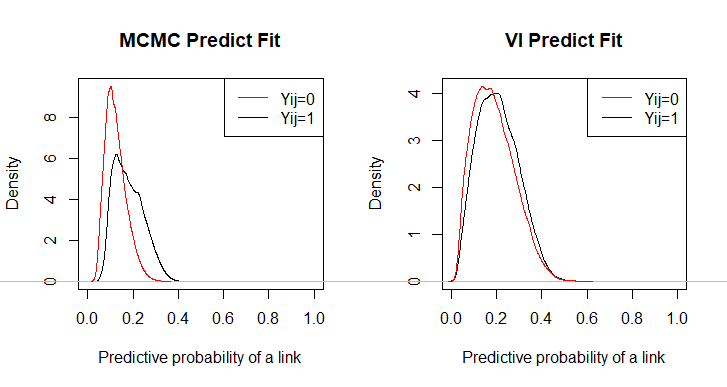}
  \caption{ Smoothed density plots of the posterior predictive probability of a link for the Florentine family dataset. The left and right plots were obtained using MCMC and BBVI, respectively. Each figure distinguishes between observations with $y_{ij}=1$ (black) and $y_{ij}=0$ (red). We observe that the faster variational method performs similarly to the MCMC method.  }
  \label{fig:spp}
\end{figure}

For this dataset we estimate our model using our proposed MCMC scheme with $100,000$ iterations and variational inference procedure with $1,000$ iterations and $S=20$. The mean-field variational family is chosen to be
\begin{align}
    q = \mathcal{N}(\alpha | \tilde{m}, \tilde{\sigma})  \prod_{i=1}^N \mbox{vMF}(z_i | \tilde{z}_i, \tilde{k}),
\end{align}
where $\mbox{vMF}(\cdot | \tilde{z}, \tilde{k})$ denotes the von-Mises-Fisher distribution (see Section \ref{sec:ell_geom}). The execution times for the MCMC and BBVI procedures were 1,227 seconds and 376 seconds, respectively. Further implementation details for each scheme with spherical geometry are given in Appendix \ref{app:mh_details} and \ref{app:spherical_bbvi_derivations}. \\

Figures \ref{fig:convergence_florentine} and \ref{fig:florentine_parameters} demonstrate convergence for each of the estimation schemes. We note that, in contrast with \cite{hoff2002}, we update each latent position individually. However, we observe that the initialisation performed well and allowed a reduction in the burn-in period for the MCMC scheme. As in \cite{hoff2002}, we also observe high variability in the results due to the random walk and tuning parameters of the MCMC. Overall, we see close correspondence between the two schemes in terms of estimates of $\alpha$ (see Table \ref{table:alpha_florentine}) and the latent positions (see Figure \ref{fig:slatent}). We note also that, since spherical space is bounded, we observe that the traceplot of the log-likelihood in Figure \ref{fig:convergence_florentine} has an upper limit. Finally, Figure \ref{fig:spp} shows that both schemes also behave similarly in terms of their predictive probability distributions. This suggests that little information is lost when we rely on the approximate variational inference procedure. 

\subsection{Karate Club}

We now consider Zachary's Karate club network \cite{zachary1977} which describes social ties among $N=34$ members of a karate club. For this dataset, $y_{ij}=1$ indicates that individuals $i$ and $j$ interacted. Similarly to Section \ref{sec:florentine_family_spherical}, we use goodness-of-fit of MDS estimates to determine an appropriate underlying geometry. Using the \text{hydra} R package \cite{hydrapackage}, we find that this data is well-suited to a model with a hyperbolic latent space. Furthermore, this data exhibit tree-like structures which further suggests a hyperbolic latent space is appropriate (see \cite{smith2019}). \\

\begin{center}

\begin{figure}[H]
  \centering
  \hspace*{-2cm}
  \includegraphics[scale=0.75]{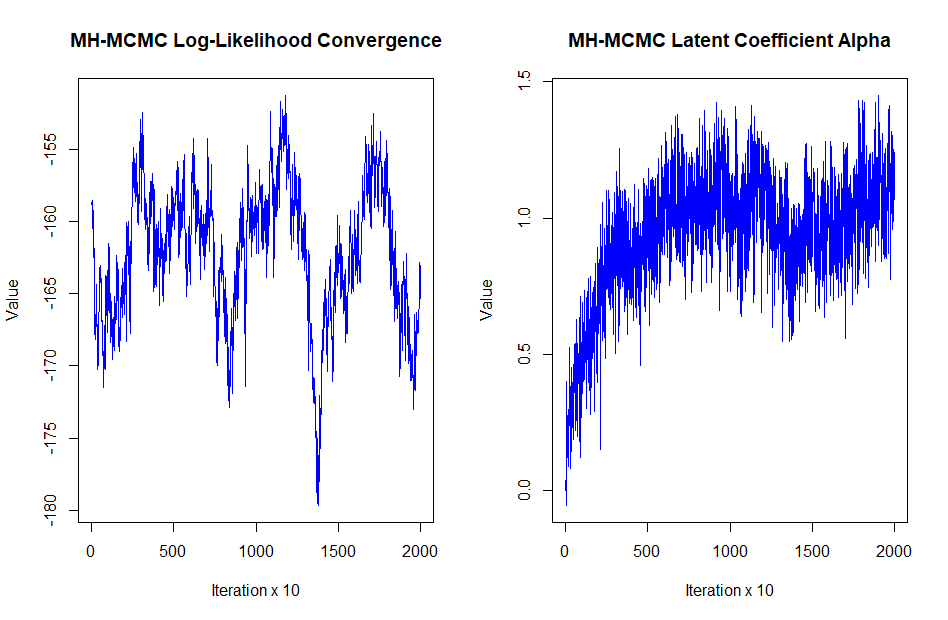}
  \caption{Thinned MCMC traceplots of log-likelihood (left) and $\alpha$ (right) for the Karate Club dataset modelled with hyperbolic latent coordinates. For each chain we plot 2000 equidistant samples from the 20000 posterior samples. }
  \label{fig:convergence_karate}
\end{figure}
\end{center}

\begin{center}

\begin{figure}[H]
  \centering
    \hspace*{-2cm}
  \includegraphics[scale=0.75]{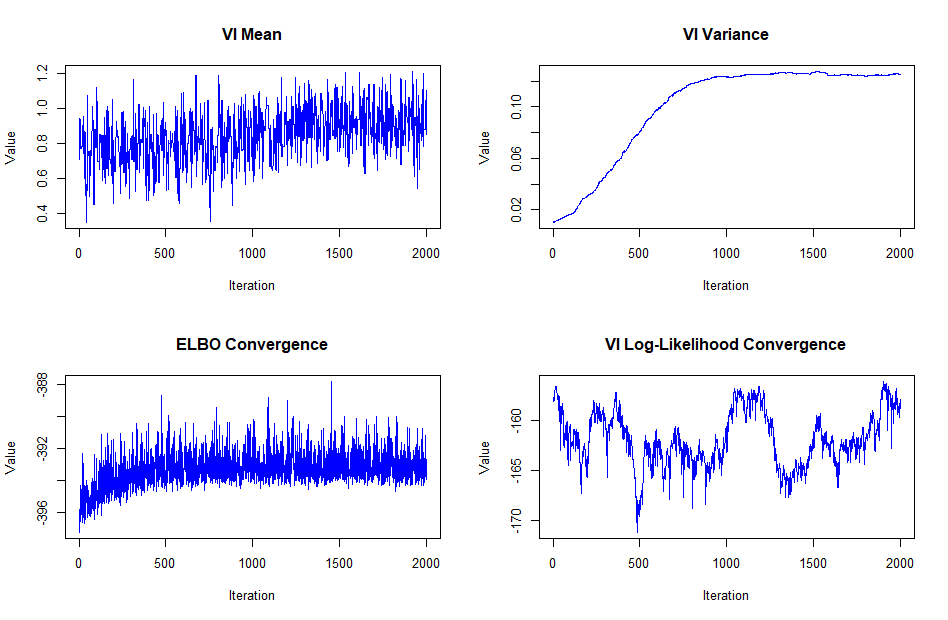}
  \caption{
  Summary of BBVI variational parameters for the Karate club network modelled with hyperbolic latent coordinates. BBVI is implemented with 2000 iterations amd $S=20$. Top, left to right: variational mean $\tilde{m}$ and variance $\tilde{\sigma}$ of $\alpha$. Bottom, left to right: ELBO and log-likelihood showing convergence for BBVI. 
  }
  \label{fig:karate_parameters}
\end{figure}
\end{center}

Similarly to Section \ref{sec:florentine_family_spherical}, we estimate our model using MCMC with $100,000$ iterations and BBVI with $1,000$ iterations and $S=20$. The mean-field variational family is chosen to be
\begin{align}
    q = \mathcal{N}( \alpha | \tilde{m}, \tilde{\sigma} ) \prod_{i=1}^N \mathcal{N}_{\mathbb{H}} ( z_i | \tilde{z}_i, \tilde{s}_i ) \label{eq:q_hyperbolic}
\end{align}
where $\mathcal{N}_{\mathbb{H}}$ denotes a hyperbolic Normal (see Section \ref{sec:hyp_geom}). The execution times for the MCMC and BBVI procedures were 2030 seconds and 594 seconds, respectively. Further implementation details for each scheme with hyperbolic geometry are given in Appendix \ref{app:mh_details} and \ref{app:hyperbolic_bbvi_details}, and we highlight here that we parameterise $\tilde{z}_i \in \mathbb{H}^2$ in polar coordinates with $\tilde{r}_i \in (-1,1)$ and $\tilde{\phi}_i \in [0, 2 \pi)$. \\

\begin{figure}[H]
  \centering
  \hspace*{-3cm}
  \includegraphics[scale=0.8]{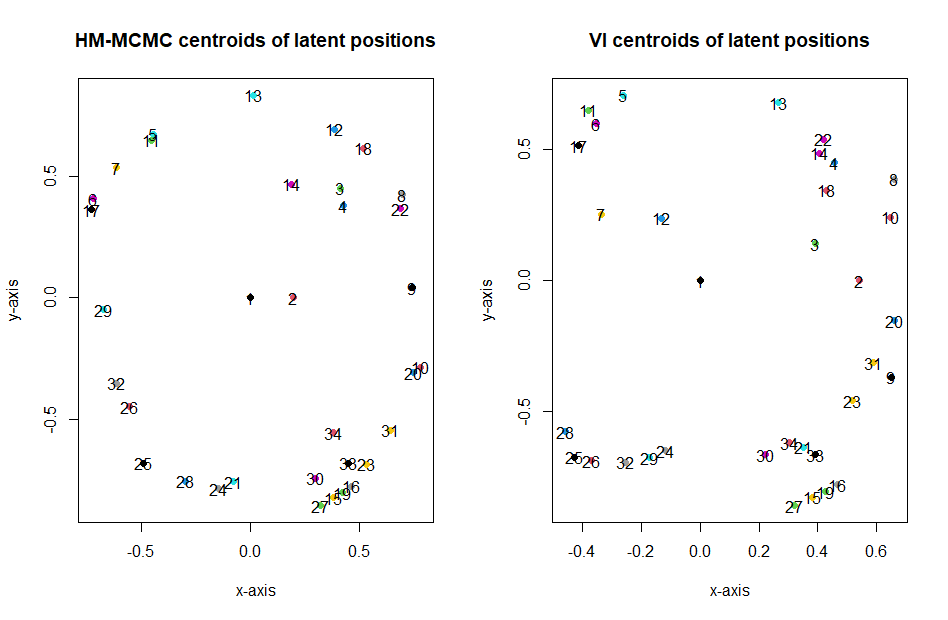}
  \caption{
  Summary of estimates of hyperbolic latent positions $\bm{Z}$ for the Karate club dataset. Left: Posterior means of $\bm{Z}$ obtained via MCMC. Right: variational mean parameter estimates $\tilde{\bm{Z}}$ obtained via BBVI. Note that we report the Fr\'{e}chet mean since this respects the spherical geometry. We also maintained the same anchor coordinates for each procedure so that the corresponding estimates are comparable.
  }
  \label{fig:hlatent}
\end{figure}

\begin{figure}[H]
  \centering
  \hspace*{-1cm}
  \includegraphics[scale=0.8]{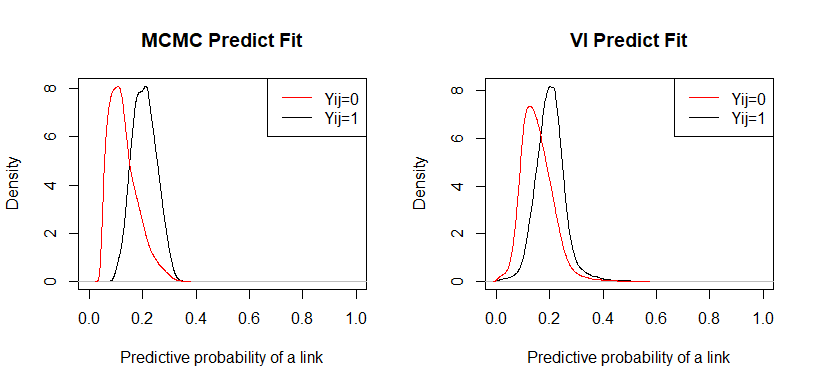}
  \caption{Smoothed density plots of the posterior predictive probability of a link for the Karate club dataset. The left and right plots were obtained using MCMC and BBVI, respectively. Each figure distinguishes between observations with $y_{ij}=1$ (black) and $y_{ij}=0$ (red). We observe that the faster variational method performs similarly to the MCMC method.}
  \label{fig:hpp}
\end{figure}

Figures \ref{fig:convergence_karate} and \ref{fig:karate_parameters} demonstrate convergence for each of the estimation schemes. Similarly to the spherical case, we note that each latent position is updated individually (in contrast to \cite{hoff2002}) and that the initialisation performed well and therefore allowed a reduction in the burn-in period for the MCMC scheme. Overall, we see that both procedures behave similarly in terms of estimates for $\alpha$ (see Table \ref{table:alpha_karate}) and the latent positions (see Figure \ref{fig:hlatent}). Finally, Figure \ref{fig:hpp} also demonstrates similar behaviour in terms of the posterior predictive distributions, suggesting that little information is lost in the approximates variational scheme. 

\begin{table}
\centering

\fbox{%
\begin{tabular}{| l  l  l l|}
\hline
Dataset  & Nodes & MCMC estimate & BBVI estimates
 \\
\hline            
Karate Club &34&  $\hat{\alpha}=1.018$&$(\tilde{m},\tilde{\sigma})$=(1.192,0.1002)\\
\hline
\end{tabular}}
\caption{Estimation of the base-rate parameter $\alpha$ for the Karate club dataset when the latent geometry is assumed to be hyperbolic. This table reports the posterior mean of $\alpha$ for the last 2000 MCMC estimations and the estimates of the variational parameters for $\alpha$.}
\label{table:alpha_karate}
\end{table}

\section{Discussion}\label{sec:disc}

In this work we have considered latent space network models with non-Euclidean geometries. Our contributions address practical considerations associated with the estimation and implementation of these models, and complement the existing literature on this topic. In particular, we have characterised the non-identifiability of the latent positions when they are assumed to lie in hyperbolic and spherical geometries, and presented procedures for Bayesian estimation via MCMC and variational methods. \\

To address the non-identifiability of the latent coordinates we propose a procedure in which a subset of the latent positions are constrained. This approach is similar to Bookstein coordinates from shape theory (see \cite{dryden1998}) and avoids the additional computation required by Procrustes for each iteration in our estimation procedures. We have presented the details of our approach for each geometry when $d=2$ and, whilst it is possible to extend this to the case where $d > 2$, this requires further analytical computations. In contrast to this, we acknowledge that relying on Procrustes makes the extension to higher-dimensions more straightforward. \\

By deriving models analogous to that of \cite{hoff2002}, it is possible to consider modelling extensions similar to those proposed in the Euclidean setting in each of our geometries of interest. This suggests many avenues for future research, such as modelling community structures \cite{handcock2007model, fosdick2016}, multiview networks  \cite{salter2017latent, dangelo2020} and dynamic networks \cite{sewell2015latent}. It is also possible to explore how alternative link functions, such as the Gaussian suggested in \cite{rastelli2016properties}, affect the properties of the generative model in conjunction with the choice of underlying geometry. Furthermore, this article assumes throughout that the latent geometry and dimension are known. Recent work has suggested procedures and heuristics to inform this choice (see \cite{lubold2020identifying, smith2019}), and future work may include adapting our approach to additionally estimate the curvature of the latent space. Finally, whilst we have considered variational methods in this work to address scalability issues, it would be interesting to also consider likelihood approximations proposed in the Euclidean setting (see \cite{raftery2012}, \cite{rastelli2018}, \cite{spencer2020}) in each non-Euclidean geometry. However, we stress that adapting this methodology to the non-Euclidean setting may not be straightforward. 


\printbibliography

\newpage

\appendix

\section{Additional details for non-Euclidean Normal distributions}

\subsection{Sampling}\label{app:sample_normal}

\begin{algorithm}[t]
  \caption{Sample from Riemannian Hyperbolic Normal $(d=2)$ (\cite{mathieu2019})} \label{alg:sampl_rhn}
  \begin{algorithmic} 
    \STATE \textbf{Input} Number of samples $n$, mean $\mu \in \mathbb{B}^2$ and dispersion $\sigma \in \mathbb{R}_{>0}$
    \STATE Calculate the envelope as
    \begin{align}
      M = \dfrac{ \gamma(2) \sigma^2 }{ Z(\sigma) } \exp \left( \dfrac{(\sigma + 1 )^2}{2} \right)
    \end{align}
    \STATE \textbf{While} $\# \mbox{samples} < n$
    \STATE \hspace{.5cm} 1) Sample a proposal location $\bm{a}$ as $\bm{a} = \sqrt{u} ( \cos \zeta, \sin \zeta )$, where $u \sim U([0,1])$ and $\zeta \sim U([0, 2 \pi])$. \\
    \STATE \hspace{1cm} (this samples uniformly within the disk $\mathcal{S}^1$)
    \STATE \hspace{.5cm} 2) Sample a proposal magnitude $r \sim \Gamma( 2, \sigma) $
    \STATE \hspace{.5cm} 3) Accept $z = \exp_{\mu} \left( \dfrac{r}{\lambda_{\mu}} \bm{a} \right)$ as a sample from the Riemannian hyperbolic Normal with probability
    \begin{align}
      AR = \dfrac{ \rho(r) }{ M  p(r | 2, \sigma^2) } = \dfrac{\dfrac{1}{Z(\sigma)} e^{ - \dfrac{r^2}{2 \sigma^2}} \sinh (r ) }{ \dfrac{M}{\Gamma(2) \sigma^2} r e^{-r/ \sigma} }
    \end{align}
\end{algorithmic}
\end{algorithm}

We cannot sample directly from the hyperbolic Normal distribution \eqref{eq:hyp_gauss}. Instead, we rely on a rejection sampler and the details of this are given in Algorithm \ref{alg:sampl_rhn}. For the spherical Normal \eqref{eq:vmf}, we use an existing rejection sampler implementation in the \texttt{movMF} R package \cite{movmf}. \\

\subsection{Unimodality}

\subsection*{Hyperbolic case}

\begin{align*}
f(z\mid\mu,\sigma^2)=\frac{1}{Z_R}\text{exp}\left(-\frac{d(z-\mu)}{2\sigma^2}\right)\\
\end{align*}

Taking the log we get
\begin{align*}
\text{log}(1)-\text{log}\left(\sqrt{2\pi\sigma^2}\right)-\frac{d_{\mathcal{H}}(z-\mu)}{2\sigma^2}
\end{align*}
Setting the derivative equal to zero we get
\begin{align*}
\frac{d}{dz}\text{log }f(z\mid\mu,\sigma^2)=-\frac{d_{\mathcal{H}}^{'}(x-\mu)}{\sigma^2}=0  \\
\Rightarrow \frac{1}{\sigma_2^2}\operatorname{arcosh}(1+2{\frac {\lVert \mu-z\rVert ^{2}}{(1-\lVert \mu \rVert ^{2})(1-\lVert z\rVert ^{2})}})\times -\frac{1}{\sqrt{1-(1+2{\frac {\lVert \mu-z\rVert ^{2}}{(1-\lVert \mu \rVert ^{2})(1-\lVert z\rVert ^{2})}})^2}} \times\\ [4{\frac {\lVert \mu-z\rVert ^{2}}{(1-\lVert \mu \rVert ^{2})(1-\lVert z\rVert ^{2})}}-4{\frac {\lVert \mu-z\rVert ^{2}}{(1-\lVert \mu \rVert ^{2})^{-2}(1-\lVert z\rVert ^{2})}} \mid \mid z \mid \mid]=0
\end{align*}
We need to prove that the above expression has only one root, which is $z=\mu$. If $z \neq \mu$ all of the three terms can not be 0 (the second one can never be zero).

\subsection*{Elliptic case}
The von-Mises Fisher distribution is unimodal for $\kappa >0$, and is uniform on the sphere for $\kappa =0$.

\section{Configuration space calculations}
\label{app:config_calc}

\subsection{Hyperbolic case}\label{app:config_hyp}

Here we determine the values of $\alpha$ and $\beta$ which satisfy
\begin{align}
    z_{i_1}^* &= 0 = h(z_{i_1}) \\
    z_{i_2}^* &= a = h(z_{i_2})
\end{align}

Since the isometry \eqref{eq:poincare_isom} preserves distances, we have
\begin{align}
    d(z_{i_1}^*, z_{i_2}^*) = d(0, a) &= \mbox{arccosh} \left(1 + 2 \dfrac{a^2}{1 - a^2} \right) = d(z_{i_1}, z_{i_2}) \\
    \Rightarrow a &= \sqrt{ \dfrac{\mbox{cosh} d(z_{i_1}, z_{i_2}) - 1}{1 + \mbox{cosh}d(z_{i_1}, z_{i_2})} }.
\end{align}

Then we obtain
\begin{align}
  h(z_{i_1}) = 0 &= \beta \dfrac{z_{i_1} - \alpha}{ \bar{\alpha} z_{i_1}- 1} \hspace{.5cm} \Rightarrow \hspace{.5cm} \alpha = z_{i_1} \\
  h(z_{i_2}) = a &= \beta \dfrac{z_{i_2} - z_{i_1}}{ \bar{z}_{i_1} z_{i_2} - 1} \hspace{.5cm} \Rightarrow \hspace{.5cm} \beta = a \left( \dfrac{\bar{z}_{i_1} z_{i_2} - 1}{z_{i_2} - z_{i_1}} \right)
\end{align}

\subsection{Spherical case}\label{app:config_sphere}

Orientation-preserving isometries in the sphere are given by compistions of the following 3-d rotation matrices.
\begin{align}
R_{z_1, \theta_{1}} = \left[
\begin{matrix}
1 & 0 & 0 \\
0 & \cos \theta_{1} & - \sin \theta_{1} \\
0 & \sin \theta_{1} & \cos \theta_{1}
\end{matrix}
\right], \hspace{.25cm}
R_{z_2, \theta_{2}} = \left[
\begin{matrix}
\cos \theta_{2} & 0 & \sin \theta_{2} \\
0 & 1 & 0 \\
- \sin \theta_{2} & 0 & \cos \theta_{2}
\end{matrix}
\right], \hspace{.25cm}
R_{z_3, \theta_{3}} = \left[
\begin{matrix}
\cos \theta_{3} & - \sin \theta_{3} & 0 \\
\sin \theta_{3} & \cos \theta_{3} & 0 \\
0 & 0 & 1
\end{matrix}
\right]. \label{eq:3d_rot_mats}
\end{align}

We take the first two anchor coordinates to be
\begin{align}
  \bm{z}_{i_1}^* = (0,0,1) \hspace{.5cm} \mbox{ and } \hspace{.5cm} \bm{z}_{i_2}^* = (a,0,b),
\end{align}
where $0<a<1$. Since the isometry preserves distances, we obtain
\begin{align}
  d_{\mathbb{S}} (\bm{z}_{i_1}, \bm{z}_{i_2} ) &=  d_{\mathbb{S}} (z_{i_1}^*, z_{i_2}^* ) = \cos^{-1} ( b) \\
  &\Rightarrow b = \cos \left( d_{\mathbb{S}} (\bm{z}_{i_1}, \bm{z}_{i_2} ) \right)
\end{align}
and, since $\bm{z}_{i_2} \in \mathbb{S}$, we have $a = \sqrt{1 - b^2}$. We take the positive root to ensure $a >0$. \\

To derive expressions for $\theta_{1}, \theta_{2}$, we consider
\begin{align}
R_{z_2, \theta_{2}} R_{z_1, \theta_{1}} = 
\left[ 
    \begin{matrix}
\cos \theta_{2} & \sin \theta_{2} \sin \theta_{1} & \sin \theta_{2} \cos \theta_{1} \\
0 & \cos \theta_{1} & - \sin \theta_{1} \\
- \sin \theta_{2} & \cos \theta_{2} \sin \theta_{1} & \cos \theta_{2} \cos \theta_{1}
\end{matrix}
\right] \label{eq:rot_xy}
\end{align}
Since $\bm{z}_{i_1}^* = (0,0,1)$, this coordinate will not be affected by the final rotation in the $z_3$ axis. By looking at the first and second rows of \eqref{eq:rot_xy}, we obtain
\begin{align}
z_{i_1, 2} \cos \theta_{1} - z_{i_1, 1} \sin \theta_{1} &= 0 \\
z_{i_1, 1} \cos \theta_{2} + z_{i_1, 2} \sin \theta_{2} \sin \theta_{1} + z_{i_1,3} \sin \theta_{2} \cos \theta_{1} &= 0
\end{align}

Now, we obtain an expression for $\theta_{3}$ by looking at
\begin{align}
R_{z_3, \theta_{3}} R_{z_2, \theta_{2}} R_{z_1, \theta_{1}} = 
\left[
    \begin{matrix}
\cos \theta_{3} \cos \theta_{2} & \cos \theta_{3} \sin \theta_{2} \sin \theta_{1} - \sin \theta_{3} \cos \theta_{1} & \cos \theta_{3} \sin \theta_{2} \cos \theta_{1} + \sin \theta_{3} \sin \theta_{2} \\
\sin \theta_{3} \cos \theta_{2} & \sin \theta_{3} \sin \theta_{2} \sin \theta_{1} + \cos \theta_{3} \cos \theta_{1} & \sin \theta_{1} \sin \theta_{2} \cos \theta_{1} - \cos \theta_{3} \sin \theta_{1} \\
- \sin \theta_{2} & \cos \theta_{2} \sin \theta_{1} & \cos \theta_{2} \cos \theta_{1}
\end{matrix}
\right] \label{eq:rot_xyz}
\end{align}

Then, from $(a,0,b)^T = R_{z_3, \theta_{3}} R_{z_2, \theta_{2}} R_{z_1, \theta_{1}} \bm{z}_{i_2}^T $, and looking at the second row of \eqref{eq:rot_xyz} we find
\begin{align}
  z_{i_2, 1} \sin \theta_{3} \cos \theta_{2} +  z_{i_2, 2}\left( \sin \theta_{3} \sin \theta_{2} \sin \theta_{1} + \cos \theta_{3} \cos \theta_{1} \right) + z_{i_2, 3} \left( \sin \theta_{1} \sin \theta_{2} \cos \theta_{1} - \cos \theta_{3} \sin \theta_{1} \right) = 0.
\end{align}

Finally, we need to apply a reflection in the plane over the first and third dimension to ensure $z_{i_3}^*$ has a positive second element.

\section{Details for MCMC estimation}\label{app:mh_details}

\subsection{Initialisation}

In order to apply either estimation procedure, we need to determine initial values of $\bm{Z}^{(0)}$ and $\alpha^{(0)}$. Following from the latent space network modelling literature, we initialise the latent coordinates using multidimensional scaling (MDS). Traditional MDS takes as input a symmetric matrix of distances and returns Euclidean coordinates with those corresponding distances, where the dimension of the coordinates is specified by the user. \\

Since we do not have the distances in the respective geometry, we use the graph distance as a proxy for this. Given this, we can use generalisations of MDS to non-Euclidean geometries. In the hyperbolic case, we rely on \texttt{hydra} R package (\cite{hydrapackage}) which implements the embedding method of \cite{kellerressel2020}. In the spherical case \texttt{smacof} R package (\cite{smacofpackage}) which implements MDS on a sphere using majorisation (see \cite{deleeuw2009}).\\

Given initial values of $\bm{Z}^{(0)}$, we can then determine $\alpha^{(0)}$ by a simple grid search. We opt to take the value of $\alpha^{(0)}$ which maximises the likelihood $p( \mathcal{Y} | \bm{Z}^{(0)}, \alpha)$.

\subsection{Priors for hyperbolic case}

Due to the nature of the distance, we use non-informative uniform prior for $\mu$ in Poincare disk and $\sigma$ in euclidean space. To distribute $N$ points uniformly at
random in a hyperbolic circle of radius $R$ where $R=1$, angular coordinates
$\theta \in [0, 2\pi]$ are sampled with the uniform density $\rho(\theta)=1/(2\pi)$, and
radial coordinates $r \in [0, R]$ are sampled with the exponential density $\rho(r)=\frac{sinh(r)}{coshR - 1} \approx e^{r-R}$.
\begin{align*}
    p(\mu) \propto e^{r-R}\\
    p(\sigma) \propto 1
\end{align*}

\subsection{Priors for spherical case}

We use the joint prior from \cite{straub2017}. In this reference paper, a marginal prior is presented, as well.

\section{Details for variational inference estimation}\label{app:vi_details}

{\subsection{Black Box VI algorithm}}

\begin{algorithm}[t] 
\begin{algorithmic}
\STATE Input: data $x$, joint distribution $p$, mean field variational family $q$. 
\STATE Initialize $\lambda$ randomly, $t=1$.
\STATE Repeat
\STATE \hspace{.5cm} Draw $S$ samples from the variational approximation
\STATE \hspace{.5cm} For $s=1$ to $S$ do
\STATE \hspace{1cm} $z[s] \sim q$
\STATE \hspace{.5cm} end for
\STATE \hspace{.5cm} For $i = 1$ to $n$ do
\STATE \hspace{1cm}For $s = 1$ to $S$ do
\STATE \hspace{1.5cm}$f_i[s]=\triangledown_{\lambda_i} \log(q[z(s)\mid \lambda_i) [\log(p(x,z[s])-\log(q(z[s]\mid \lambda_i)]$
\STATE \hspace{1.5cm}$h_i[s]=\triangledown_{\lambda_i} \log(q[z(s)\mid \lambda_i)$
\STATE \hspace{1cm}end for
\STATE \hspace{1cm}$\hat{\alpha}_i^{*}=\frac{\sum_{d=1}^{n_i}\hat{Cov}(f_i^d,h_i^d)}{\hat{Var}(h_i^d)}$
\STATE \hspace{1cm}$\hat{\triangledown}_{\lambda_i}\mathcal{L}\overset{\Delta}{=}\frac{1}{S}\sum_{s=1}^Sf_i[s]-\hat{\alpha}_i^{*}h_i[s]$
\STATE \hspace{.5cm}end for
\STATE \hspace{.5cm}$\rho$ = $t^{th}$ value of a Robbins Monro sequence
\STATE \hspace{.5cm}$\lambda= \lambda+\rho \triangledown_{\lambda} \mathcal{L}$
\STATE \hspace{.5cm}$t=t+1$
\STATE until change of $\lambda$ is less than 0.01.
\end{algorithmic}
\caption{Black Box Variational Inference (Algorithm 2 of \cite{ranganath2014})} \label{alg:bbvi}
\end{algorithm}

Algorithm \ref{alg:bbvi} outlines the details of the the BBVI sampler. In this description, we aim to target the joint distribution $p(x,z)$ where $x$ represent the data and $z$ represent the latent parameters, with the mean-field variational family $q$ parameterised by $\lambda$. In our setting we have $x = \mathcal{Y}, z = \{\bm{Z}, \alpha, \theta_z\}$ and $\lambda = \{\tilde{m}, \tilde{\sigma},\tilde{\theta}_z\}$. Following \cite{ranganath2014} we calculate the scaling parameter $\rho$ using rmsprop (\cite{duchi2011}). We rely on this procedure to estimate the hyperbolic and spherical model variations, and calculations required for this are presented in the following subsections.

\subsubsection{Hyperbolic case}
\label{app:hyperbolic_bbvi_details}

We take a mean field variational family
\begin{align}
  q = \prod_{i=1}^N q(z_i | \tilde{z}_i, \tilde{s}_i ) q( \alpha | \tilde{m}, \tilde{\sigma} )
\end{align}
where $z_i$ follows a hyperbolic Gaussian (see \eqref{eq:hyp_gauss}) and $\alpha$ follows a Normal distribution so that
\begin{align}
  \log q( \alpha | \tilde{m}, \tilde{\sigma}) &= -\dfrac{1}{2} \log 2 \pi - \log \tilde{\sigma} - \dfrac{1}{2} \left(\dfrac{\alpha-\tilde{m}}{\tilde{\sigma}} \right)^2 \\
  \log q( z_i | \tilde{z}_i, \tilde{s}_i ) &= - \log \left( 2 \pi \dfrac{\sqrt{\pi}}{2} \right) - \log \tilde{s}_i - \dfrac{ \tilde{s}_i^2}{2} - \log \mbox{erf} \left( \dfrac{\tilde{s}_i}{\sqrt{2}} \right) - \dfrac{d^2_{\mathcal{P}}(z_i, \tilde{z}_i) }{2 \tilde{s}_i^2}
\end{align}
Comparing to \eqref{eq:q}, we have $\tilde{\theta}_z = \left( \{\tilde{z}_i\}_{i\in [N]}, \{ \tilde{s}_i \}_{i \in [N]} \right) $. Now, for each parameter, we require an expression of the gradient of $\log q$. We have
\begin{align}
      \dfrac{\partial}{\partial \tilde{m}} \log q( \alpha | \tilde{m}, \tilde{\sigma}^2) &= \dfrac{\alpha - \tilde{m}}{\tilde{\sigma}^2}  \\
  \dfrac{\partial}{\partial \tilde{\sigma}} \log q( \alpha | \tilde{m}, \tilde{\sigma}^2) &= - \dfrac{1}{\tilde{\sigma}} + \dfrac{(\alpha - \tilde{m})^2}{\tilde{\sigma}^3}  \\
  \dfrac{\partial}{\partial \tilde{z}_i } \log q(z_i | \tilde{z}_i, \tilde{s}_i) &= - \dfrac{1}{2 \tilde{s}^2 } \dfrac{\partial}{\partial \tilde{z}_i } d_{\mathcal{P}}^2( z_i, \tilde{z}_i )  = - \dfrac{  d_{\mathcal{P}}( z_i, \tilde{z}_i ) }{\tilde{s}_i^2} { \dfrac{\partial}{\partial \tilde{z}_i } d_{\mathcal{P}}( z_i, \tilde{z}_i ) } \\
  \dfrac{\partial}{\partial \tilde{s}_i } \log q(z_i | \tilde{z}_i, \tilde{s}_i) &= - \dfrac{1}{\tilde{s}_i } - \tilde{s}_i - { \dfrac{\partial}{\partial \tilde{s}_i } \log \mbox{erf} \left( \dfrac{\tilde{s}_i}{\sqrt{2}} \right) } + \dfrac{d^2_{\mathcal{P}}(z_i, \tilde{z}_i) }{\tilde{s}_i^3}
\end{align}
where
\begin{align}
{ \dfrac{\partial}{\partial \tilde{z}_i } d_{\mathcal{P}}(\tilde{z}_i, z_i) } &=   \dfrac{\partial}{\partial \tilde{z}_i }  \cosh^{-1} \left(1 + 2 \dfrac{\| \tilde{z}_i-z_i \|^2}{ (1 - \|\tilde{z}_i\|^2)(1 - \|z_i\|^2)} \right) \\
                                                                                                &=   \dfrac{\partial}{\partial y }  \cosh^{-1} (y) \dfrac{ d y}{ d \tilde{z}_i }  \\
                                                                                                &= \dfrac{1}{\sqrt{y^2 - 1}}  \dfrac{\partial}{\partial \tilde{z}_i }  \left(1 + 2 \dfrac{\| \tilde{z}_i-z_i \|^2}{ (1 - \|\tilde{z}_i\|^2)(1 - \|z_i\|^2)} \right) \\
                                                                                                &= \dfrac{1}{\sqrt{y^2 - 1}} \dfrac{2}{1 - \|z_i\|^2} \dfrac{\partial}{\partial \tilde{z}_i } \left(  \dfrac{\| \tilde{z}_i-z_i \|^2}{ (1 - \|\tilde{z}_i\|^2)} \right) \\
                                                                                                &= \dfrac{1}{\sqrt{y^2 - 1}} \left( \dfrac{2}{1 - \|z_i\|^2} \right) \left(  \dfrac{ 2( \tilde{z}_i-z_i)}{ 1 - \|\tilde{z}_i\|^2 } + \dfrac{2 \tilde{z}_i \| \tilde{z}_i - z_i\|}{ (1 - \|\tilde{z}_i\|^2)^2 } \right)  \\
{ \dfrac{\partial}{\partial \tilde{s}_i } \log \mbox{erf} \left( \dfrac{\tilde{s}_i}{\sqrt{2}} \right) } &= \dfrac{ \dfrac{\partial}{\partial \tilde{s}_i } \mbox{erf} \left( \dfrac{\tilde{s}_i}{\sqrt{2}} \right) }{ \mbox{erf} \left( \dfrac{\tilde{s}_i}{\sqrt{2}} \right) } = \dfrac{2 e^{- \tilde{s}_i^2 / 2}}{ \sqrt{2 \pi} \mbox{erf} \left( \tilde{s}_i / \sqrt{2} \right) }
\end{align}

In BBVI, the variational parameters are updated on $\mathbb{R}$. This requires us to update some of the variational parameters on different scales. $\tilde{m}$ is unconstrained, but we update $\tilde{s}_i$ and $\tilde{\sigma}$ on the log scale as $\tilde{s}_i^* = \log \tilde{s}_i$ and $\tilde{\sigma}^* = \log \tilde{\sigma}$. Additionally, we parameterise $\tilde{z}_i$ as $   \tilde{z}_i = (\tilde{r}_i \cos \tilde{\upvarphi}_i, \tilde{r}_i \sin \tilde{\upvarphi}_i) $, where $\tilde{r}_i \in [0,1]$ and $\tilde{\upvarphi}_i \in \mathbb{R}$. $\tilde{\upvarphi}_i$ is updated as unconstrained and we update $\tilde{r}_i^* \in \mathbb{R}$ where
\begin{align}
  \tilde{r}_i = \dfrac{1}{ 1+ e^{- \tilde{r}_i^*}}
\end{align}

The required gradients are then given by applying the chain rule.

\subsubsection{Spherical case}
\label{app:spherical_bbvi_derivations} 

We take a mean field variational family
\begin{align}
  q = p(\alpha | \tilde{m}, \tilde{\sigma}) \prod_{i=1}^N q( z_i | \tilde{z}_i, \tilde{k}_i )
\end{align}
where $p(\alpha | \tilde{m}, \tilde{\sigma}) = \mathcal{N}(\alpha | \tilde{m}, \tilde{\sigma}) $ and $ q( z_i | \tilde{z}_i, \tilde{k} ) = \mbox{vMF}(z_i | \tilde{z}_i, \tilde{k})$ (see \eqref{eq:vmf}) so that
\begin{align}
  \log q( z_i | \tilde{z}_i, \tilde{\kappa}_i) &= \log(\tilde{\kappa}_i) - \log(2\pi) - \log\left( e^{\tilde{\kappa}_i} - e^{-\tilde{\kappa}_i}\right) + \tilde{\kappa}_i \tilde{z}_i^T z_i \\
  \log q( \alpha | \tilde{m}, \tilde{\sigma} ) &= -\dfrac{1}{2} \log 2 \pi - \log \tilde{\sigma} - \dfrac{1}{2} \left(\dfrac{\alpha-\tilde{m}}{\tilde{\sigma}} \right)^2 
\end{align}
Following the note \cite{straub2017}, we use the more numerically stable expression for the von-Mises-Fisher given by
\begin{align}
  \log q( z_i | \tilde{z}_i, \tilde{\kappa}_i) = \log( \tilde{\kappa}_i ) - \log(2 \pi) - \log \left( 1 - e^{ - 2 \tilde{\kappa}_i } \right) + \tilde{\kappa}_i ( \tilde{z}_i^T z_i - 1).
\end{align}

As in the hyperbolic case, we require an expression for the gradient of $\log q$ for each variational parameter. We have
\begin{align}
    \dfrac{\partial}{\partial \tilde{m}} \log q( \alpha | \tilde{m}, \tilde{\sigma}^2) &= \dfrac{\alpha - \tilde{m}}{\tilde{\sigma}^2} \label{eq:tm_upd} \\
  \dfrac{\partial}{\partial \tilde{\sigma}} \log q( \alpha | \tilde{m}, \tilde{\sigma}^2) &= - \dfrac{1}{\tilde{\sigma}} + \dfrac{(\alpha - \tilde{m})^2}{\tilde{\sigma}^3}  \label{eq:tsig_upd} \\
  \dfrac{\partial}{\partial \tilde{\kappa}_i} \log q( z_i | \tilde{z}_i, \tilde{\kappa}_i ) &= \dfrac{1}{\tilde{\kappa}_i} - \dfrac{e^{\tilde{\kappa}_i} + e^{-\tilde{\kappa}_i}}{e^{\tilde{\kappa}_i} - e^{-\tilde{\kappa}_i}} + \tilde{z}_i^T z_i = \dfrac{1}{\tilde{\kappa}_i} + (\tilde{z}_i^T z_i - 1) - \dfrac{2 e^{-2 \tilde{\kappa}_i}}{1 - e^{- \tilde{\kappa}_i}} \\
  \dfrac{\partial}{\partial \tilde{z}_i} \log q( z_i | \tilde{z}_i, \tilde{\kappa}_i ) &= \tilde{\kappa}_i z_i
\end{align}
where the second expression for the gradient with respect to $\tilde{\kappa}_i$ is more numerically stable. \\

We update $\tilde{m}$ as unconstrained, and take the variances on the log scale as $\tilde{\kappa}_i^* = \log \tilde{\kappa}_i$ and $\tilde{\sigma}^* = \log \tilde{\sigma}$. To update $\tilde{z}_i$, we transform to polar coordinates so that

\begin{align}
  \tilde{z}_i = (\tilde{u}_{i1}, \tilde{u}_{i2}, \tilde{u}_{i3}) = ( \cos \tilde{\phi}_i \sin \tilde{\omega}_i, \sin \tilde{\phi}_i \sin \tilde{\omega}_i , \cos \tilde{\omega}_i )
\end{align}
where $\tilde{\omega}_i \in [0, \pi]$ and $\tilde{\phi}_i \in [0, 2 \pi)$. Since trigonometric functions are periodic, we do not need to further constrain the angles in the updates.
Then the updates then are given by the chain rule 
\begin{align}
   \dfrac{\partial}{\partial \tilde{\sigma}^*} \log q( \alpha | \tilde{m}, \tilde{\sigma}^2) &= \dfrac{\partial}{\partial \tilde{\sigma}} \log q( \alpha | \tilde{m}, \tilde{\sigma}^2) \dfrac{\partial \tilde{\sigma}}{\partial \tilde{\sigma^*}} = \dfrac{\partial}{\partial \tilde{\sigma}} \log q( \alpha | \tilde{m}, \tilde{\sigma}^2) \times \tilde{\sigma} \\
     \dfrac{\partial}{\partial \tilde{\kappa}_i^*} \log q( z_i | \tilde{z}_i, \tilde{\kappa}_i ) &= \dfrac{\partial}{\partial \tilde{\kappa}_i} \log q( z_i | \tilde{z}_i, \tilde{\kappa}_i ) \times \kappa_i \\
  \dfrac{\partial}{\partial \tilde{\omega_i}} \log q( z_i | \tilde{z}_i, \tilde{\kappa}_i ) &= \tilde{\kappa}_i \dfrac{\partial \tilde{z}_i^T}{\partial \tilde{\omega_i}} z_i  = \tilde{\kappa}_i ( \cos \tilde{\phi}_i \cos \tilde{\omega}_i, \sin \tilde{\phi}_i \cos \tilde{\omega}_i , -\sin \tilde{\omega}_i )^T z_i \\
    \dfrac{\partial}{\partial \tilde{\phi_i}} \log q( z_i | \tilde{z}_i, \tilde{\kappa}_i ) &= \tilde{\kappa}_i \dfrac{\partial \tilde{z}_i^T}{\partial \tilde{\phi_i}} z_i  = \tilde{\kappa}_i (  - \sin \tilde{\phi}_i \sin \tilde{\omega}_i, \cos \tilde{\phi}_i \sin \tilde{\omega}_i , 0 )^T z_i
\end{align}

\section{Lattice Geometry}
\label{sec:latt_geom}

Lattice geometry does not present any peculiarity that inherently describes well known properties of networks. We refer, describe and analyze lattice geometry in advance for future research reasons. For example, an interesting topic is the analysis of mixing values which, as we mention, differ depending on the design of the grid. Latent space models using Euclidean, instead of lattice, geometry in terms of the speed of convergence and the accuracy of the results is superior. Next, we provide evidence that support the statements above.

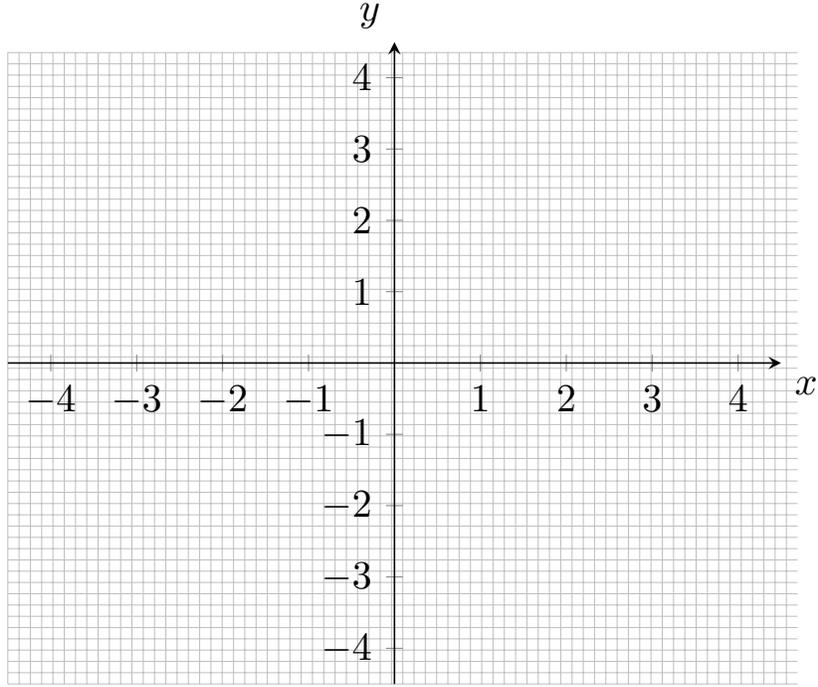
\begin{figure}
\begin{center}
\begin{tikzpicture}[scale=1.5]
\begin{scope}
\draw [very thin, lightgray,step=.1] (0,0) grid (7,5.6);
\begin{axis}[
  axis x line=center,
  axis y line=center,
  xtick={-4,-3,...,4},
  ytick={-4,-3,...,4},
  xlabel={$x$},
  ylabel={$y$},
  xlabel style={below right},
  ylabel style={above left},
  xmin=-4.5,
  xmax=4.5,
  ymin=-4.5,
  ymax=4.5]
\end{axis}
\end{scope}
\end{tikzpicture}
\end{center}
\caption{Lattice geometry points. They discretize euclidean geometry and every point is a potential latent position for the network. In our framework, between points (0,0) and $(0,a)$ in x-axis $m$ points exist.}
\end{figure}

\subsection{Geometry}

A lattice is a discrete additive subgroup of $\mathbb{R}^m$ and a set of linearly independent vectors that generates a lattice, denoted $B = {b_1,\dots,b_n} \subset \mathbb{R}^m$ for integers $m \geq n \geq$ 1,  is called a basis. The lattice generated by the basis $B$ is given by
\begin{equation}
\mathbb{L} = \mathcal{L}(B) = \left\{ Bz=\sum_{i=1}^n z_ib_i : z \in \mathbb{Z}^n \right\}. \label{eq:lb}
\end{equation}
We say that the rank of this lattice is $n$ and its dimension is $m$. \\

The definition in \eqref{eq:lb} is very general and, for the remainder of this article, we restrict our attention to the $d$-dimensional lattice given by a product over $d$ copies of $\mathbb{Z}$. We denote this as
\begin{equation}
\mathbb{L}_{\mathbb{Z}}^d = \left\{ \sum_{i=1}^n a_i e_i : a \in \mathbb{Z}^d \right\}, \label{eq:l_z}
\end{equation}
where $e_i$ is standard basis vector of $\mathbb{R}^d$ whose $i^{th}$ entry is 1 and all other entries are 0. For example, when $d=2$, \eqref{eq:l_z} denotes all integer combinations of the basis $B = \{(0,1), (1,0)\}$. Note that $\mathbb{L}_{\mathbb{Z}}^d$ is a special case of \eqref{eq:lb} in which the dimension and rank are equal. \\

In $\mathbb{L}_{\mathbb{Z}}^2$, we determine the distance between two points $z_1 = (z_{11}, z_{12})$ and $z_2 = (z_{21}, z_{22})$ using
\begin{align}
  d_{\mathbb{L}_{\mathbb{Z}}} = | z_{21} - z_{11} | + | z_{22} - z_{12} |. \label{eq:lattice_dist}
\end{align}
This distance is also referred to as the Manhattan or taxicab distance metric.\\

To describe a Normal distribution on $\mathbb{L}_{\mathbb{Z}}^d$ we follow \cite{roy2013high}. Other discrete normal distributions could have been used (\cite{roy2003,agostini2019discrete,aggarwal2013note,chakraborty2015generating}), as well. For a random variable $X \sim \mathcal{N}(\mu, \sigma)$, a discrete Normal can be constructed by taking the following density function.
\begin{align}
    p(X = k) = \Phi \left( \dfrac{k + 1 - \mu}{\sigma}\right) - \Phi \left( \dfrac{k  - \mu}{\sigma}\right),
\end{align}
where $k \in \mathbb{Z}$. \\

We consider a discrete Normal on $\mathbb{Z}^2$ with zero mean and dispersion $\sigma \in \mathbb{R}_{>0}$, where we note that $\sigma$ is close but not equal to the standard deviation. To define this let
\begin{equation}
S(k) = \sum_{k=-\infty}^{\infty} e^{-k^2/(2\sigma^2)}= 1 + 2\sum_{k=1}^{\infty}e^{-k^2/(2\sigma^2)}
\end{equation}
and let $E$ be the random variable on $\mathbb{L}$, that follows a discrete Normal with zero mean and dispersion $\sigma$ such that, for $k \in \mathbb{Z}^2$, 
\begin{equation}
    Pr(E = k) = \rho_{\sigma}(k) = \frac{1}{S(k)}e^{-k^2/(2\sigma^2)}.
\end{equation}
Given this, a discrete Normal distribution with mean $c \in \mathbb{Z}$ can now be expressed as  $\rho_{\sigma,c}(x) = \frac{1}{S} \exp(-(x -c)/(2 \sigma^2))$. The 2-dimensional lattice is given by the two dimensional grid of $\mathbb{Z}^2$.\\

{\bf \Large {Priors for lattice case}}\\

The improper priors for $c$ and $\sigma$ are:
\begin{align*}
    p(c)=p(c_1,c_2)=p(c_1)p(c_2) \propto 1\\
    p(\sigma) \propto \frac{1}{\sigma}
\end{align*}

{\bf \Large {Unimodality}}\\

\begin{align*}
p(z\mid c,\sigma)=\frac{e^{-\frac{\mid\mid z_i-\mu\mid\mid^2}{2\sigma^2}}}{(\sum_{j=-\infty}^{\infty}  e^{{-\frac{1}{2\sigma^2}\mid\mid z_j-\mu \mid\mid^2}})}
\end{align*}
The above sequence $(z_1, \ldots, z_n)$ is unimodal with mode $\mu$ because it enjoys the property that:
\begin{align*}
     z_1 < \cdots < z_{\mu-1} < z_{\mu} > z_{\mu+1} > \cdots > z_n
\end{align*}
where $1 \leq \mu < n$. So, we have a maximum at $z_i=\mu$.

\subsection{Non-identifiability}\label{app:config_lattice}

In $\mathbb{R}^d$, the isometries can be expressed as combinations of rotations, translations and reflections. Since the lattice $\mathbb{L}_{\mathbb{Z}}$ defined in \eqref{eq:l_z}, can be thought of as a restriction of $\mathbb{R}^d$, is follows that the isometries can be viewed as a restriction of the Euclidean isometries as follows.
\begin{enumerate}
\item Translations $T(z)$ can be written as $T(z) = z + a$, where $a \in \mathbb{L}_{\mathbb{Z}}$.
\item Reflections are restricted to be along a line parallel to either $(0,1)$ or $(1,0)$.
\end{enumerate}

Note that these transformations do not commute, and the restrictions may be modified so that they are appropriate for different choices of lattice geometry \eqref{eq:lb}. \\

The easiest way to get a lattice is to discretize $\mathbb{R}^2$. The latent positions of the network are then associated with the lattice points. Before dicretization, a get rid of the identiability issues by applying the isometries or translation and reflection, sending two points to x-axis and a third one below x-axis. In $\mathbb{R}^2$, we set the first two anchor points $z^*_{i_1} = (z^*_{i_1,1}, z^*_{i_1,2})$ and $z^*_{i_2} = (z^*_{i_2,1}, z^*_{i_2,2})$ to be (0, 0) and $(0, a)$, respectively, where $a$ is specified so that $d(z^*_{i_1}, z^*_{i_2})=d(z_{i_1}, z_{i_2})$. We define an isometry as translation, followed by a reflection on the x-axis (expressed, here, for convenient purposes as a rotation across the origin), so that
\begin{align}
z^* = R(z - b)=\begin{bmatrix}
\cos(\phi) & -\sin(\phi)\\
 \sin(\phi) & \cos(\phi)
\end{bmatrix}
\left(z -
\begin{bmatrix}
z_{i_1,1}^*+z_{i_2,1}^*\\
z_{i_1,2}^*+z_{i_2,2}^*
\end{bmatrix}
\right) \label{eq:isom_latt}
\end{align}
where $\phi = \arctan \left( {\frac{z^*_{i_2,2} - z^*_{i_1,2}}{z^*_{i_2,1} - z^*_{i_1,1}}} \right)$. Finally, we take the third anchor coordinate to be $z_{i_3}^* = (b,c)$ where $c>0$. If after the transformation \eqref{eq:isom_latt} has been applied we find that $z_{i_3}^*$ has a negative second element, we then apply a reflection in the axis of the first dimension.\\

We take the anchor points to be
\begin{align}
  \bm{z}_{i_1}^* = (0,0) \hspace{.5cm} \mbox{ and } \hspace{.5cm} \bm{z}_{i_2}^* = (0,a),
\end{align}
where $a \in \mathbb{Z}_{+}$. The desired isometry can be expressed as $\bm{z} \mapsto R( \bm{z} - b)$ where $b \in \mathbb{L}_{\mathbb{Z}}$ and
\begin{align}
R = 
\left[ 
  \begin{matrix}
    \cos \phi & - \sin \phi \\
    \sin \phi & \cos \phi
\end{matrix}
    \right]
\end{align}

Since $\bm{z}_{i_1} \mapsto (0,0)$, it immediately follows that $b = \bm{z}_{i_1}$. Then we require
\begin{align}
  R ( \bm{z}_{i_2} - \bm{z}_{i_1} ) = (0,a).
\end{align}
By reading the first row of this expression, we obtain
\begin{align}
  \cos \phi (z_{i_2,1} - z_{i_2,1} ) = \sin \phi (z_{i_2,2} - z_{i_2,2} )
\end{align}
and the value of $\phi$, which is connected with $a$ follows immediately.

As we mentioned the transformations in the lattice, namely translation and reflection, do not commute. In $\mathbb{Z}^2$, we set the first two anchor points $z^*_{i_1} = (z^*_{i_1,1}, z^*_{i_1,2})$ and $z^*_{i_2} = (z^*_{i_2,1}, z^*_{i_2,2})$ to be $(a, b)$ and $(c, d)$, respectively, where the points are specified so that $d(z^*_{i_1}, z^*_{i_2})=d(z_{i_1}, z_{i_2})$. $z^*_{i_2}$ is fixed along the line of $(c-a,b-d)$. After the transformations has been applied we force a third point $z_{i_3}^*$ to have elements at the same side of the line $(c-a,b-d)$, so that the reflection in the line is taken into account.\\

Discretization of $\mathbb{R}$, with $m$ increments between 0 and $z^{*}_{i_i}$, is then conducted. Every point in $\mathbb{R}^2$ is rounded to the closest point in $\mathbb{Z}^2$ based on minimum the euclidean distance. For different values of $m$, different results are produced through the Bayesian algorithms. Further comments and results regarding discrete distributions, random walks and lattice geometry are presented in the appendix \ref{sec:latt_geom}.

\subsection{Lattice case - Stein Variational Gradient Descent}

We use the same notation of \cite{han2020stein}: Given a discrete distribution $p_{*}$, there are many different continuous parameterizations. Because exact samples of $p_{c}$ yield exact samples of $p_{*}$ following the definition, we should prefer to choose continuous parameterizations whose $p_{c}$ is easy to sample from using continuous inference method. So, as in \cite{han2020stein}, the discrete Stein gradient decent becomes a weighted Stein gradient decent as in the \ref{alg:stein_discrete}.

\begin{algorithm}[H] 
\begin{algorithmic}
\STATE Goal: Approximate a given distribution $p_{*}(z)$ (input) on a finite discrete set $Z$.
\begin{itemize}
\item Decide a base distribution $p_0(x)$ on $\mathbb{R}^d$(such
as a Gaussian distribution for each dimension), and a map $\Gamma: \mathbb{R}^d \rightarrow Z$ which partitions $p_0$ evenly. Construct a piecewise continuous distribution $p_c$: $p_c(x) \propto p_0(x)p_{*}(\Gamma(x))$.
\item Construct a differentiable surrogate of $p_c(x)$, for
example, by $\rho(x) \propto p_0(x)$.
\item  Run gradient-free SVGD on $p_c$ with differentiable surrogate $\rho$: starting from an initial $\{x_i\}_n^{i=1}$ and
repeat $x_i \leftarrow x_i+\frac{\epsilon}{\sum_i w_i} \sum_{j=1}^n w_j(\triangledown \rho(x_j)k(x_j,x_i)+\triangledown_{x_j}k(x_j,x_i))$, where $w_j = \rho(x_j)/p_c(x_j)$, and $k(x, x^{'}$ is a positive
definite kernel.
\item  Calculate $z_i = \Gamma(x_i)$ and output sample $\{z_i\}^n_{i=1}$ for approximating discrete target distribution $p_{*}(z)$.
\end{itemize}
\end{algorithmic}
\caption{GF-SVGD on Discrete Distributions}
\label{alg:stein_discrete}
\end{algorithm}

For the experiments, we used 100 particles to approximate the distributions. \\

The base function $p_0(x)$ is the p.d.f. of the bivariate standard Gaussian distribution. Applying the map $z = \Gamma(x) = sign(x)$ for discrete two dimensional $x$, the transformed piecewise continuous target is $p_c(x) \propto p_0(x)p^{\mbox{*}}(sign(x))$.\\

To construct the differentiable surrogate $\rho$ in the algorithm \ref{alg:stein_discrete}, we set $\rho=p_0(x)$, which is the multiplication of normal continuous normal distributions for each dimension/parameter of interest.

For the discrete, lattice geometry we use the discrete Stein Variational inference procedure in \cite{han2020stein}. This method is based on a stochastic optimization of the variational objective where existing forms of gradient descent can not be directly applied to discrete distributions, such as in the BBVI algorithm. We use as a base geometry the 2-dimensional Gaussian distribution to approximate the piecewise Gaussian distribution, as described.

 Lattice grid on $\mathbb{Z}^2$.


\subsection{Example}

For the lattice case the initialisation is similar with \cite{hoff2002}, because of the similarity of those distributions. This is happening because it is the discrete version of initialisation in \cite{hoff2002} and we use the discredited version of euclidean MDS.

\subsection*{Sampson Network}

The purpose of this simulation is twofold. We consider Sampson's Monk dataset \cite{sampson1969crisis,handcock2008statnet} to provide evidence that the distance between each points in the lattice geometry can influence the mixing of the parameters and we present an implementation of an MCMC and VI algorithm for networks in the lattice. Sampson Monks network is an ethnographic study of community structure in a New England monastery conducted by Samuel F. Sampson. This network describes the interpersonal relations among $N$ = 18 monks, during Sampson's stay in a monastery, while a political "crisis in the cloister" resulted in the expulsion of four monks and the voluntary departure of several others.\\

\begin{figure}[H]
  \centering
    \hspace*{-2cm}
  \includegraphics[scale=0.7]{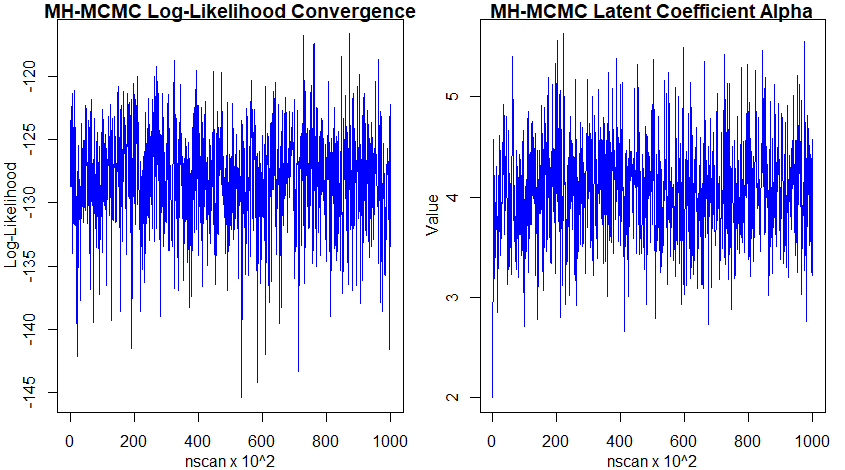}
  \caption{MCMC outcomes of \cite{hoff2002} paper for log-likelihood and value of latent coefficient.}
  \label{fig:mcmcconv}
\end{figure}

\begin{figure}[H]
  \centering
    \hspace*{-2cm}
  \includegraphics[scale=0.7]{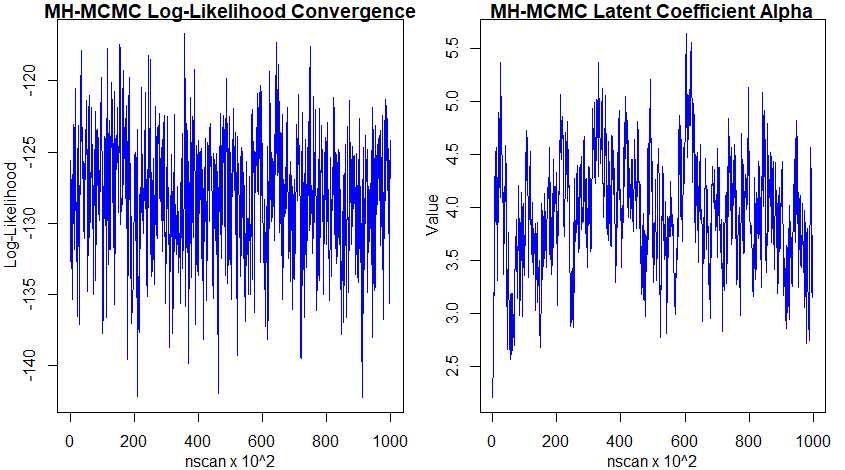}
  \caption{MCMC outcomes of lattice geometry, with $m$ increments of distance 0.001, for log-likelihood and value of latent coefficient.}
  \label{fig:mcmcconv1}
\end{figure}

We implement an MCMC with 100000 iterations, and VI Stein (\cite{han2020stein}) with 2000 iterations and 20 particles. For the MH-MCMC and the VI, with distance between each points 0.001, the execution times were 1715 seconds and 184 seconds accordingly. With figures \ref{fig:mcmcconv} and \ref{fig:mcmcconv1} we see that lattice geometry can be used to obtain similar results, as in \cite{hoff2002}. In line with \cite{hoff2002}, all latent position parameters are updated at once in each step of MCMC. Moreover, latent coefficient $a$ is forced to be positive as in \cite{hoff2002}. We, observe, that initialization worked as well as in euclidean case. As in \cite{hoff2002}, the high variability in the results, due to the random walk and tuning parameters of the MCMC, are depicted. \\

\begin{figure}[H]
  \centering
    \hspace*{-2cm}
  \includegraphics[scale=0.8]{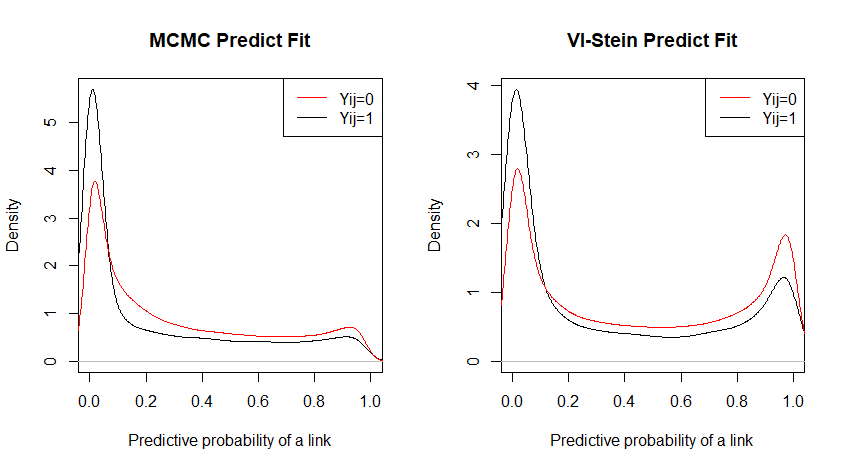}
  \caption{Comparison of two methods for Sampson Monk data set using the lattice. Left: MH-MCMC posterior predictive distribution. Right: BBVI posterior predictive distribution.  }
  \label{fig:latticepredic}
\end{figure}

 In Figure \ref{fig:latticepredic}, smoothed density plots of the posterior predictive probability of a link for the Sampson Monk
dataset is illustrated for both methods. The posterior predictive probabilities are split according to whether the data shows a link
($Y_{ij}=1$) or not ($Y_{ij}=0$). We see that the faster variational method performs almost identically to the MCMC method.

\subsection*{Mixing results}

\begin{figure}[H]
  \centering
    \hspace*{-2cm}
  \includegraphics[scale=0.8]{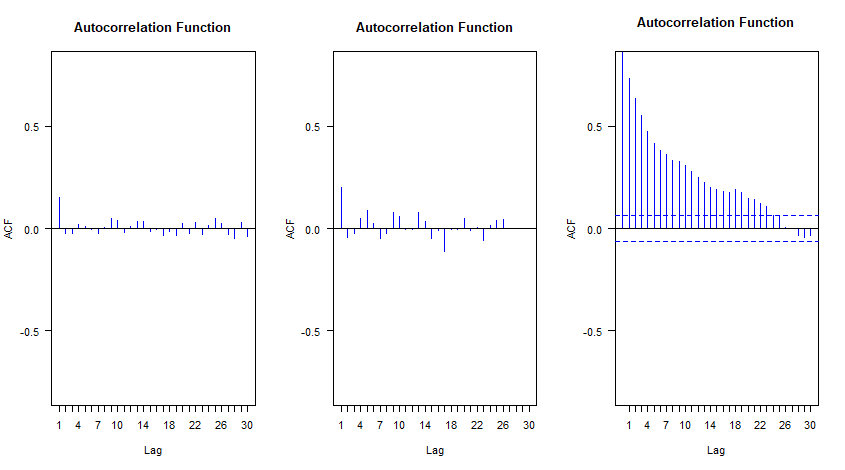}
  \caption{Autocorrelation function of latent coefficient Alpha. Left: Continuous case Middle: Discrete case with $m$ increments of 0.001 Right: Discrete case with $m$ increments of 0.1}
  \label{fig:mixing}
\end{figure}

For fast-mixing Markov chains, lag-$k$ autocorrelation values drop down to (practically) zero quickly as k increases. On the other hand, high lag-$k$ autocorrelation within chains indicate slow mixing and, usually, slow convergence. To examine whether lattice space allows us to achieve better mixing time than the euclidean case demonstrated in \cite{hoff2002} (figure \ref{fig:mixing}), we produce the autocorrelation function of latent coefficient parameter of \cite{hoff2002} and compare it to two different lattice grids. We observe first that  that under the right parametrization, lattice geometry offers as good mixing time as we can achieve in euclidean case. Though, by increasing the distance between points in the lattice, from 0.001 to 0.1, does not improve the mixing (figure \ref{fig:mixing}).\\

Further investigation about how and under which properties lattice geometry offers better mixing time in networks or hypergraphs is a topic of specific interest.

\end{document}